\documentclass[11pt,aps,prd,nofootinbib,superscriptaddress,preprintnumbers]{revtex4}
\pdfoutput=1
\usepackage{graphicx,epsfig}
\usepackage{amsmath,amsfonts,amssymb,amsthm,amscd,amsbsy,xfrac}
\usepackage{calc}
\makeatletter
\usepackage{color}
\def\met{\ensuremath{E_{\mathrm{T}}^{\mathrm{miss}}}\,} 

\def\pt{\ensuremath{p_{\mathrm{T}}}\,}

\def\mll{\ensuremath{m_{\mathrm{ll}}}\,} 
 
\def\HT{\ensuremath{H_{\mathrm{T}}}\,}

\def\GeV{\ensuremath{\mathrm{GeV}}\,}

\begin{document}
\preprint{FTUV-15-03-13}
\preprint{IFIC-15-19}
\title{METing SUSY on the $Z$ peak}

\author{G.~Barenboim}
\affiliation{Departament de F\'{\i}sica Te\`orica, Universitat de 
Val\`encia, C/Dr.~Moliner 50,  E-46100, Burjassot, Spain.}
\affiliation{IFIC, Universitat de 
Val\`encia-CSIC, Parc Cient\'{\i}fic U.V., C/Catedr\'atico Jos\'e Beltr\'an 2, E-46980, Paterna, Spain.}
\author{J.~Bernabeu}
\affiliation{Departament de F\'{\i}sica Te\`orica, Universitat de 
Val\`encia, C/Dr.~Moliner 50, E-46100, Burjassot, Spain.}
\affiliation{IFIC, Universitat de 
Val\`encia-CSIC, Parc Cient\'{\i}fic U.V., C/Catedr\'atico Jos\'e Beltr\'an 2, E-46980, Paterna, Spain.}
\author{V.A.~Mitsou}
\affiliation{IFIC, Universitat de 
Val\`encia-CSIC, Parc Cient\'{\i}fic U.V., C/Catedr\'atico Jos\'e Beltr\'an 2, E-46980, Paterna, Spain.}
\author{E.~Romero}
\affiliation{IFIC, Universitat de 
Val\`encia-CSIC, Parc Cient\'{\i}fic U.V., C/Catedr\'atico Jos\'e Beltr\'an 2, E-46980, Paterna, Spain.}
\author{O. Vives}
\affiliation{Departament de F\'{\i}sica Te\`orica, Universitat de 
Val\`encia, C/Dr.~Moliner 50, E-46100, Burjassot, Spain.}
\affiliation{IFIC, Universitat de 
Val\`encia-CSIC, Parc Cient\'{\i}fic U.V., C/Catedr\'atico Jos\'e Beltr\'an 2, E-46980, Paterna, Spain.}
\begin{abstract}
Recently the ATLAS experiment announced a 3~$\sigma$ excess at the $Z$-peak consisting of 29 pairs of leptons together with two or more jets, \met $> 225$~GeV and $\HT > 600$~GeV, to be compared with $10.6 \pm 3.2$ expected lepton pairs in the Standard Model. No excess outside the $Z$-peak was observed.
By trying to explain this signal with SUSY we find that only relatively light gluinos, $m_{\tilde g} \lesssim 1.2$~TeV, together with a heavy neutralino NLSP of $m_{\tilde \chi} \gtrsim 400$~GeV decaying predominantly to $Z$-boson plus a light gravitino, such that nearly every gluino produces at least one $Z$-boson in its decay chain, could reproduce the excess. We construct an explicit general gauge mediation model able to reproduce the observed signal overcoming all the experimental limits. Needless to say, more sophisticated models could also reproduce the signal, however, any model would have to exhibit the following features, light gluinos, or heavy particles with a strong production cross-section, producing at least one $Z$-boson in its decay chain. The implications of our findings for the Run II at LHC
with the scaling on the Z peak, as well as for the direct search of
gluinos and other SUSY particles, are pointed out.
\end{abstract}
\maketitle
\section{Introduction}

   The discovery by ATLAS \cite{Armstrong:1994it,Aad:2008zzm,Aad:2009wy} and CMS \cite{CMS:1994hea} experiments at the LHC Collider \cite{LHCStudy:1995aa,Evans:2008zzb} of a
new particle with a mass of 125 GeV \cite{Aad:2012tfa,Chatrchyan:2012ufa} and with the expected properties
of a Higgs boson has marked the programme of high-energy physics for
the next coming 
years. On one side, it is mandatory to be precise enough in the
measurements of the properties of the new scalar particle in order to definitively ascertain its nature as the Higgs boson remnant of the electroweak
symmetry breaking mechanism. On the other, its mass is still 
compatible with the requirements imposed by supersymmetry (SUSY) at the
expense of moving the SUSY scale above TeV energies. Combined with the 
current LHC constraints (although model-dependent in most cases) from 
data analyses in the first run of LHC, the search for SUSY effects is 
becoming more restrictive. Nevertheless, the discovery of SUSY would be such an
extraordinary event, not only by itself, but for solving pending 
fundamental experimental and theoretical problems in particle physics,
that an intense well-motivated experimental programme to search
for SUSY effects is of the highest interest.

        Within this scenario, the ATLAS Collaboration has recently 
presented an intriguing excess at the 3 sigma level of $e^+ e^-$
and $\mu^+ \mu^-$ pairs just at the $Z$ peak \cite{Aad:2015wqa}, accompanied by hadronic 
activity and missing transverse energy (MET). With an integrated luminosity of $20.3 ~{\rm fb}^{-1}$ at a $p-p$ CM energy of 8 TeV, the experiment observes a total of 29
pairs of electrons and muons with an invariant mass compatible with the $Z$ boson mass,
with an expected background of $10.6 \pm 3.2$ pairs. No excess over
the expected background is observed outside the $Z$ peak \footnote{A similar analysis on $Z$ plus \met has been performed by CMS \cite{Khachatryan:2015lwa}. However, among other differences, no cut on $\HT$ has been applied. No deviation from SM expectations has been observed here.}. The question
that immediately arises is whether SUSY, or some other extension of the Standard Model (SM), 
is able to explain that excess of $Z$ + MET
events taking into account the current limits on beyond the SM physics. A study in those terms within a SUSY framework is
presented in this paper.

     As we will show, the observed signal can only be explained if one has a 
large production cross-section of heavy SUSY particles (gluinos or squarks) 
whose decay chain contains about one $Z$-boson per parent
particle. If such an explanation is indeed the answer to the observed excess, 
our study points out the way to confirm it in the Run II of LHC, as well as
cosmological implications, in particular the particle content of dark
matter in the Universe.
 The resulting scheme of SUSY particle mass
hierarchy, including charginos and neutralinos, will be apparent.

\section{ATLAS excess in $l^+ l^-$ on the $Z$ peak}
\label{ATLAS_selection}

In order to motivate and clarify the assumptions that will be made in the next sections, a simplified summary of the mentioned ATLAS search is given here.

The main focus of the search in \cite{Aad:2015wqa} are the decays of squarks and gluinos with two leptons (electrons or muons), jets and \met in the final state, where the two leptons originate from a $Z$-boson.
 
In order to discriminate between SM background events and a possible signal, the following requirements are applied:
\begin{itemize}
\item At least two same-flavoured leptons with opposite electric charge are required in each event. If more than two leptons are present in the event, the two with the largest values of \pt are selected. The leading lepton, i.e. the lepton with highest \pt, must have \pt $>$ 25~\GeV, whereas the subleading lepton \pt can be as low as 10~\GeV. Their invariant mass must fall within the $Z$ boson mass window, here considered as 81 $< \mll <$ 101~\GeV. 
\item All events are further required to contain at least two jets with \pt $>$ 35 \GeV  and $ |\eta| < $ 2.5, to have \met $>$ 225 \GeV  and $\HT > $ 600 \GeV, where \HT is the \pt sum over all the jets with \pt $>$ 35 \GeV  and $ |\eta| < $ 2.5 and the two leading leptons: $H_{{\rm T}} = \sum_i p_{{\rm T}}^{{\rm jet},i} + p_{{\rm T}}^{{\rm lepton},1} + p_{{\rm T}}^{{\rm lepton},2}$. 
\item Furthermore, the azimuthal angle between each of the two leading jets and \met is required to be higher than 0.4.
\end{itemize}

A great effort has been made to accurately estimate the number of SM events that survive the previous selection. An enumeration of these expected SM processes together with some of their characteristics follows:
\begin{itemize}
\item The main background, namely $t\overline{t}$, together with $W W$, $W t$ and $Z \rightarrow \tau \tau$, which add up to $\sim60\%$ of the predicted background, have been estimated using a data-driven method that has been thoroughly cross-checked with different techniques.
\item Diboson backgrounds with real $Z$-boson production ($\sim25\%$) and ``rare top'' ($t\overline{t} + W, t\overline{t} + Z, t\overline{t} + WW$ and $t + Z$) backgrounds ($<5\%$) are estimated using MC simulation. These are subject to carefully assessed theoretical and experimental uncertainties.
\item Processes with ``fake leptons'', i.e. leptons originating from the mis-identification of a jet, ($\sim10\%$) are estimated using a data-driven method used regularly in most of ATLAS analyses.
\item Finally, particular care has been taken to suppress the $Z/\gamma^{*} + jets$ background as much as possible, given that it could mimic a possible signal (the cut in the azimuthal angle between each of the two leading jets and the \met has been applied to serve this purpose). Nonetheless, a data-driven technique has been used to estimate this small ($<1\%$) but important background.
\end{itemize}

The total number of expected SM model events passing all the requirements is 10.6 $\pm$ 3.2 and the number of observed data events is 29. This corresponds to a 3.0~$\sigma$ significance.

\section{$Z$-boson production in the MSSM}

As explained above, if the observed excess is confirmed, it
would clearly point to a new non-standard process producing additional
$Z$-bosons at LHC energies. $Z$-bosons are regularly produced in the
decay chains of most of the SM extensions. Still, this signal
would require a significant production of $Z$-bosons without
conflicting with all other experimental searches of beyond the SM
particles. In fact, using the central value for the expected
background and taking into account the $Z$-branching ratio to muon and
electron, this would imply that we have produced $273\pm 48$ additional
$Z$-bosons (with $20.3 ~{\rm fb}^{-1}$). Assuming the $Z$-bosons are
produced in the decay-chains of beyond the SM particles, $Y$, produced
in the collision, we need to produce at least $273/{\cal N}(Y\to Z)$ $Y$
particles, with ${\cal N} (Y\to Z)$ the average number of $Z$-bosons produced in the decay of a $Y$ particle. On the other hand, as we will see next, the experimental cuts used in the experiment, namely $n_{\rm jets} \geq 2$, \met $>$ 225 \GeV  and $\HT > $ 600 \GeV, define further the characteristics of the $Y$ particle and its decays.  

Now, the following question is whether it is possible to
produce these extra $Z$-bosons with the associated event-attributes 
in some SUSY extension of the SM, while
at the same time all the constraints imposed by present new-physics
searches at LHC and other experiments are satisfied. In this section, we will assume that the acceptance of the applied selection, also taking into account the reconstruction efficiencies is ideally equal to unity. In Section~\ref{GGM}, we will perform a ``realistic'' simulation in a SUSY model using Delphes \cite{deFavereau:2013fsa} to take into account the signal acceptance and detector efficiencies.

\subsection{Production cross-section of MSSM particles}
\label{crosssection}
We need to produce 273 (225 at one-$\sigma$) $Z$-bosons if we want to accommodate the observed excess in lepton-pairs. Assuming that R-parity is conserved, supersymmetric particles are produced in pairs in processes of the type $p p \to Y \bar Y$. Thus, the required cross-section for this process would be,
  \begin{equation}
\sigma( p p \to Y  \bar Y) = \frac{N_{\rm ev}/{\cal N}(Y\to Z)}{\cal L} = \frac{137 (113)/{\cal N}(Y \to Z)}{20.3~ {\rm fb}^{-1} } = \frac{6.7 (5.6)~ {\rm fb}}{{\cal N}(Y\to Z)} \,,
\label{requiredXS}
  \end{equation}
where we take into account that two $Y$ particles are produced in each event.
So, if we obtained one $Z$-boson for each $Y$-particle produced, we would need a production cross-section of $(6.7 \pm 1.1)~{\rm fb}$ at the LHC with a CM energy of 8 TeV. Then, the first thing we must do is to determine whether it is still possible to have these production cross-sections for some supersymmetric particle taking into account the constraints from direct searches at LHC. Here, we consider the production cross sections of different supersymmetric particles separately to identify the relevant processes. However, in a full simulation, as done in Section~\ref{GGM}, different sparticles contribute to the final $Z$ plus jets plus MET signal and the different contributions must be added.  

Naively, the first option to consider in a hadron collider would be strong production of squarks or gluinos (assuming they produce $Z$ bosons in their decays). However, current experimental searches of jets plus missing energy at LHC force the masses of these coloured particles to be high \cite{Khachatryan:2015pwa,Aad:2015mia,Aad:2014wea}. Nevertheless, as we will see below, in some cases we can still find cross-sections of the required size. 

\begin{center}
  \begin{figure}[h!]
  \includegraphics[scale=0.62]{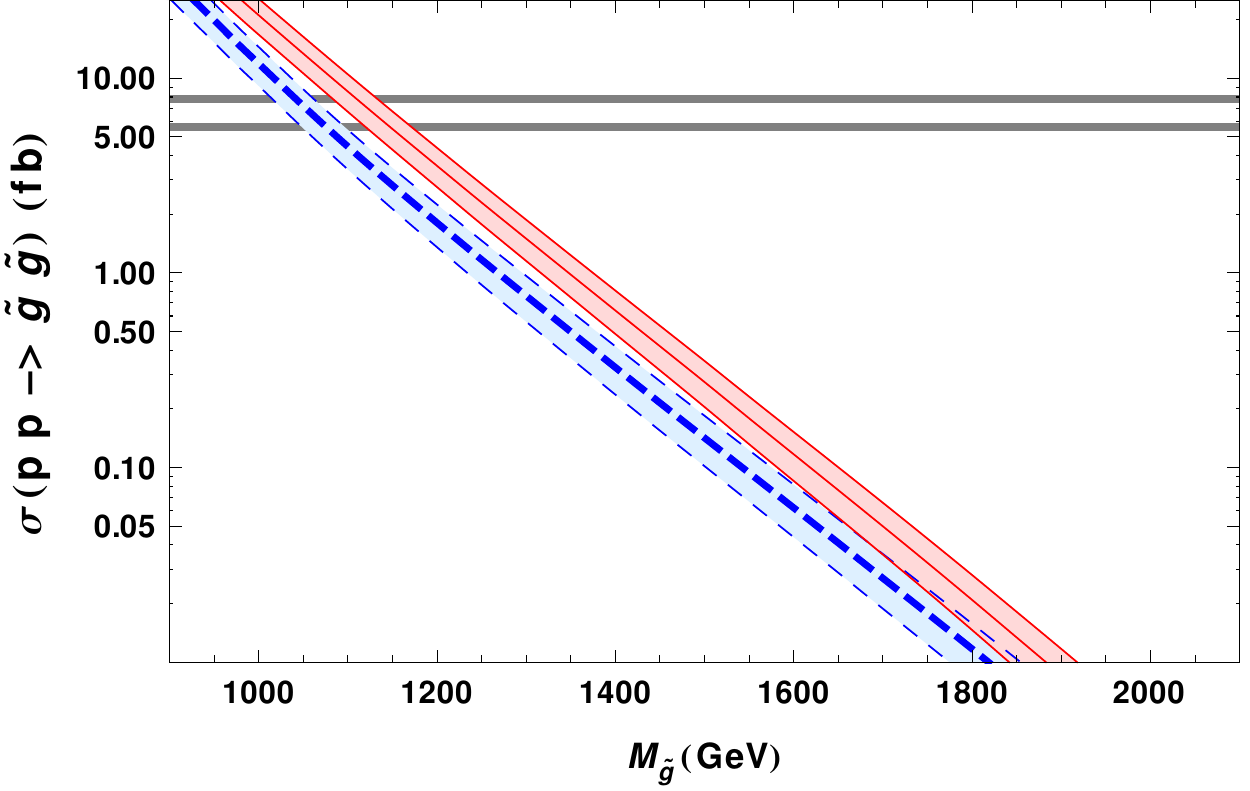} \includegraphics[scale=0.62]{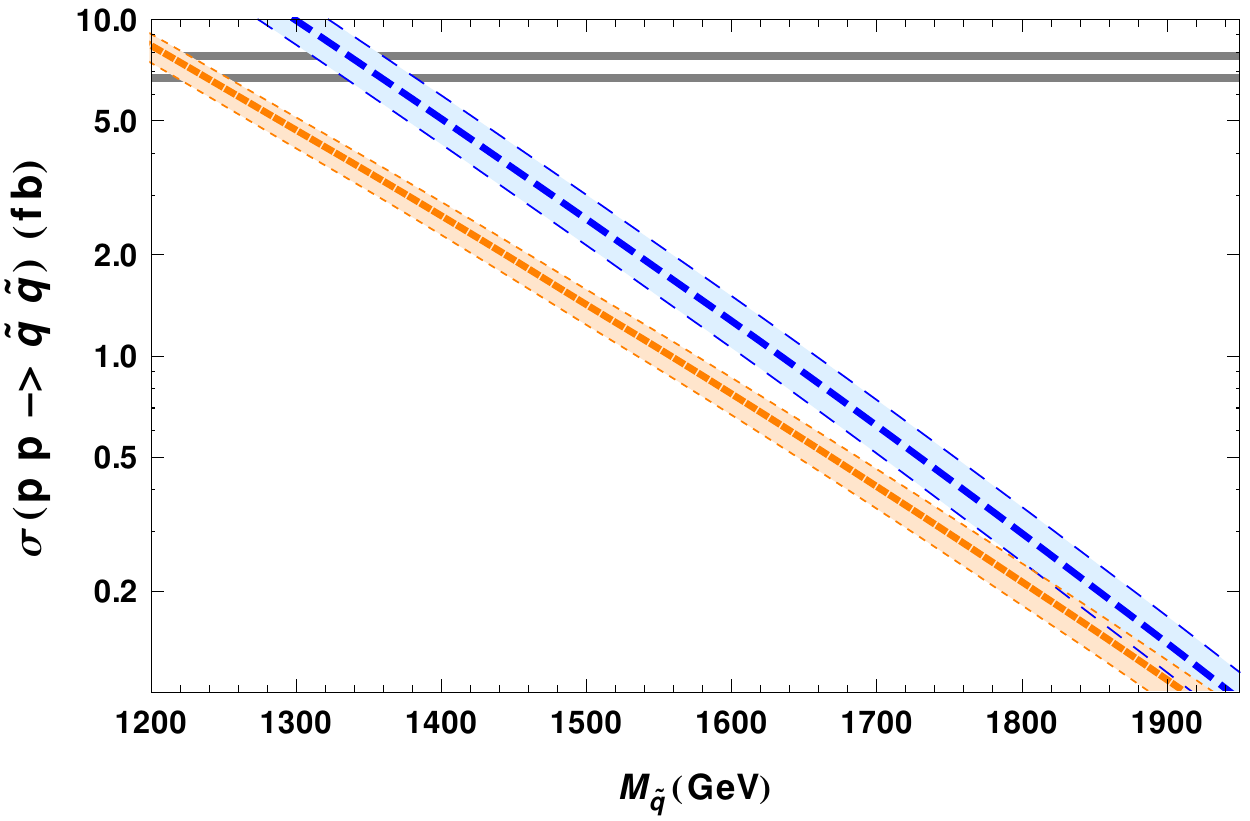}
  \caption{On the left, we show the production cross-section of gluino pairs as a function of the gluino mass for two fixed values of the first generation squark masses: 1000 GeV (blue/dashed) and decoupled squarks (red/solid). On the right, we have the  production cross-section of squarks pairs as a function of their mass with gluino masses of 1000 GeV (blue/dashed) and 2500 GeV (orange/dotted).}
\label{xsgluqq}
  \end{figure}
  \end{center}

\begin{center}
  \begin{figure}[h!]
  \includegraphics[scale=0.6]{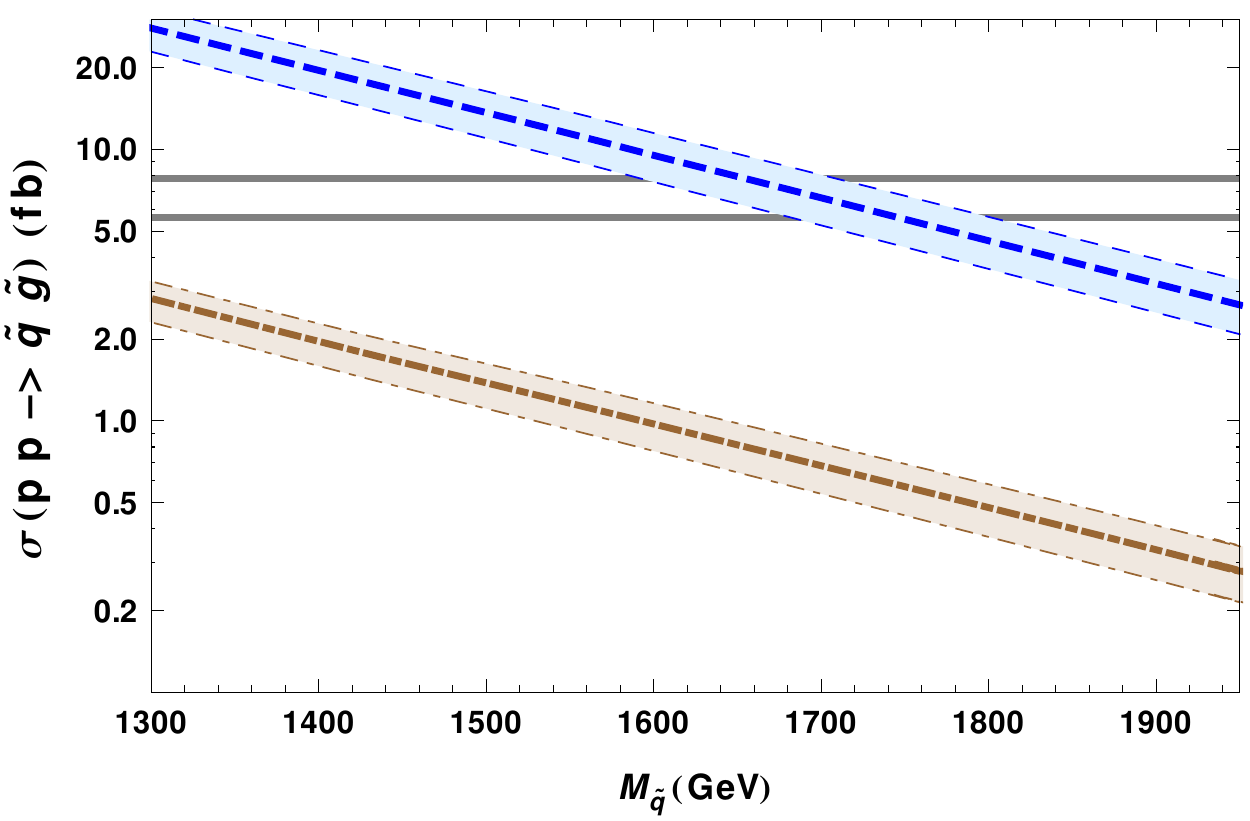} \includegraphics[scale=0.62]{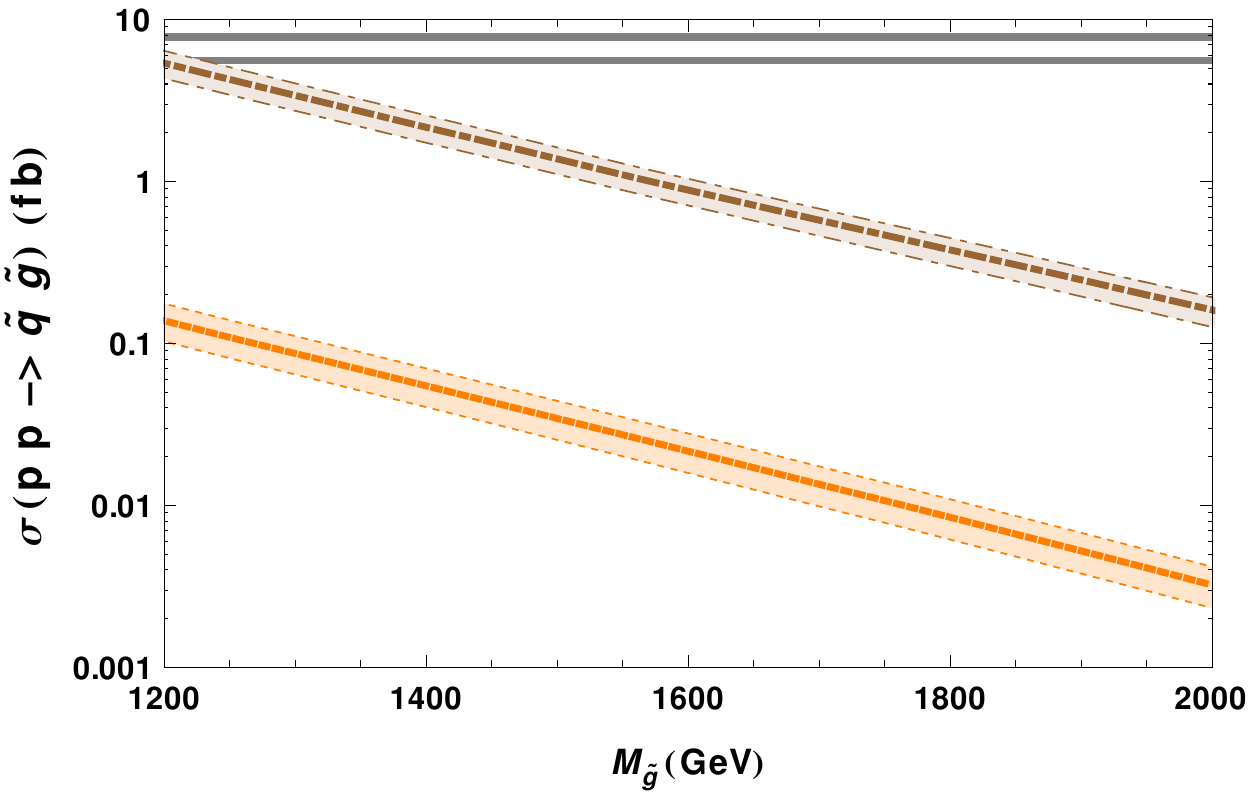}
  \caption{Production cross-section of squark-gluino as a function of the squark mass (left) or the gluino mass (right) for fixed values of the associate particle mass: 1000 GeV (blue/dashed) 1500 GeV (brown/dash-dotted) and 2500 GeV (orange/dotted).}
\label{xsqglu}
  \end{figure}
  \end{center}

Production cross-sections of gluinos and squarks depend only on their masses and are basically independent of other MSSM parameters. In the case of gluino and squarks of the first generation, the cross-section depends both on the squark and gluino masses due to t-channel contributions, but in the case of stops or sbottoms it depends only on the stop or sbottom  mass. In Fig.~\ref{xsgluqq} we present the production cross-section of gluino pairs and light-flavour squark pairs \footnote{These squark cross-sections are obtained with 5 squark flavours (mass degenerate), i.e. include contributions from sbottoms which are treated as a light flavour.} calculated at NLL+NLO with NLL-fast \cite{Beenakker:1996ch,Beenakker:1997ut,Kulesza:2008jb,Kulesza:2009kq,Beenakker:2009ha,Beenakker:2010nq,Beenakker:2011fu} as a function of the gluino or squark mass. Fig.~\ref{xsqglu} shows the squark plus gluino cross-section as a function of the squark or gluino mass with the second mass fixed at different values. The different bands in these figures correspond to the cross-section at one-$\sigma$ with fixed squark or gluino masses: the blue (dashed) band corresponds to $m_{\tilde q, \tilde g} = 1000$ GeV, the brown (dash-dotted) band to $m_{\tilde q, \tilde g} = 1500$ GeV, the orange (dotted) band has $m_{\tilde q, \tilde g} = 2500$~GeV and the red (solid) band corresponds to decoupled squarks or gluinos. In all these plots, the horizontal grey lines represent the required cross-section at one-$\sigma$ needed to reproduce the referred ATLAS excess.

As we can see in these figures, the required cross-section is reached only for light gluino and squark masses. In the case of gluino production, the needed cross-section is obtained only for $m_{\tilde g} \lesssim 1200$ GeV and favours heavy squark masses. In fact, these gluino masses are in the boundary of the allowed region obtained from jets plus missing $E_T$ searches at LHC \cite{Khachatryan:2015pwa,Aad:2015mia,Aad:2014wea} and would contribute significantly only if every $\tilde g$ produces at least a $Z$-boson in its decay. For the production of squark pairs, present limits are $m_{\tilde q} \gtrsim 1400$ GeV for heavy gluinos and  $m_{\tilde q} \gtrsim 1650$ GeV for degenerate squarks and gluinos \cite{Khachatryan:2015pwa,Aad:2015mia,Aad:2014wea}. Under these conditions, $\sigma ( p p \to \tilde q \tilde q)$ is always well-below the required cross-section, even for $m_{\tilde q} \gtrsim 1400$ GeV. Here, we do not show the cross-section $\sigma ( p p \to \tilde q \tilde q^*)$ as it is typically one order of magnitude smaller than  $\sigma ( p p \to \tilde q \tilde q)$, and thus irrelevant.

Another important process is the simultaneous production of squark and gluino shown in Fig. \ref{xsqglu}. However,  we see that  we would need  both squark and gluino to be light, which is not possible if we take into account the bounds from LHC searches \cite{Khachatryan:2015pwa,Aad:2015mia,Aad:2014wea}. 
In summary, the best option seems to be gluino pair production with $m_{\tilde g} \lesssim 1200$ GeV with relatively heavy squarks  $m_{\tilde q} \gtrsim 3000$ GeV, if we can get at least one $Z$ boson in every gluino decay.  

\begin{center}
  \begin{figure}[h!]
  \includegraphics[scale=0.8]{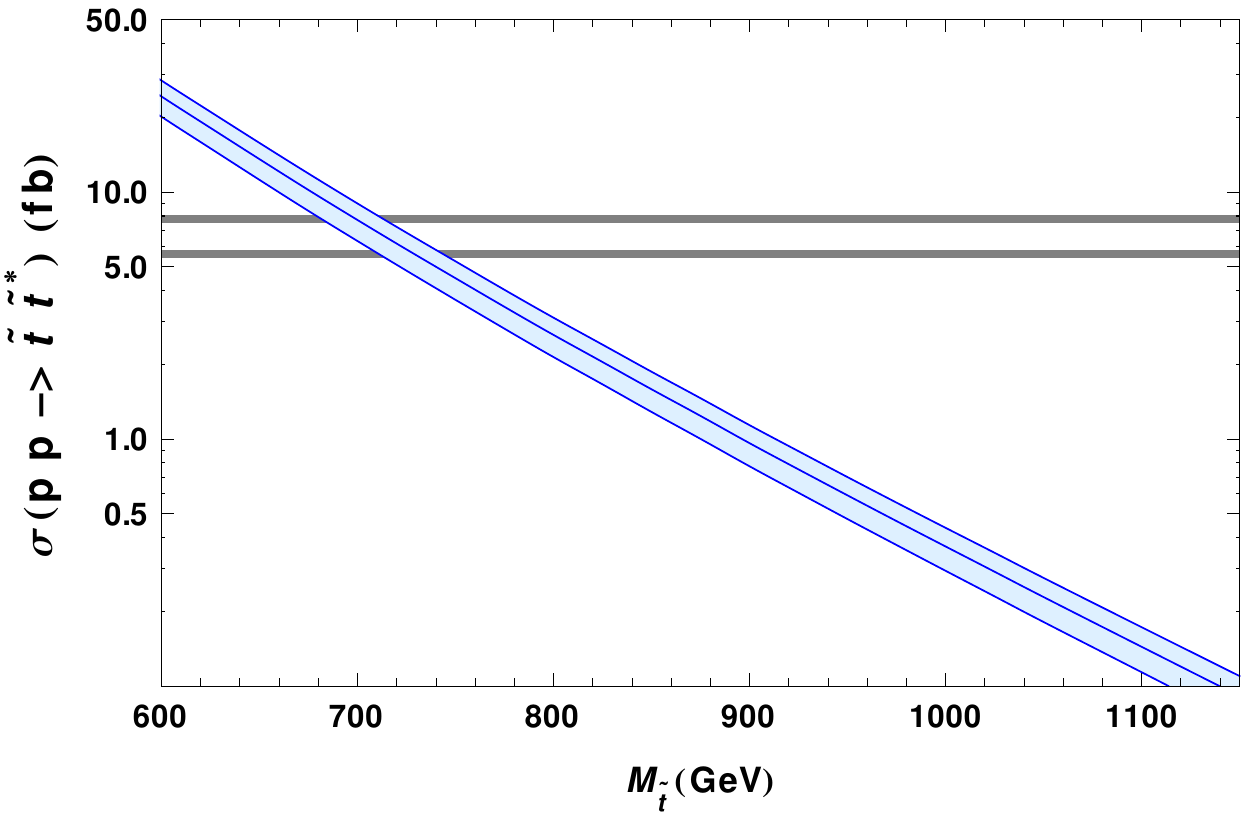}
  \caption{Production cross-section of $\tilde t \tilde t^*$ pairs as a function of the stop mass. The cross-section is basically independent of other SUSY masses. }
\label{xstt}
  \end{figure}
  \end{center}
\begin{center}
  \begin{figure}[h!]
  \includegraphics[scale=0.5]{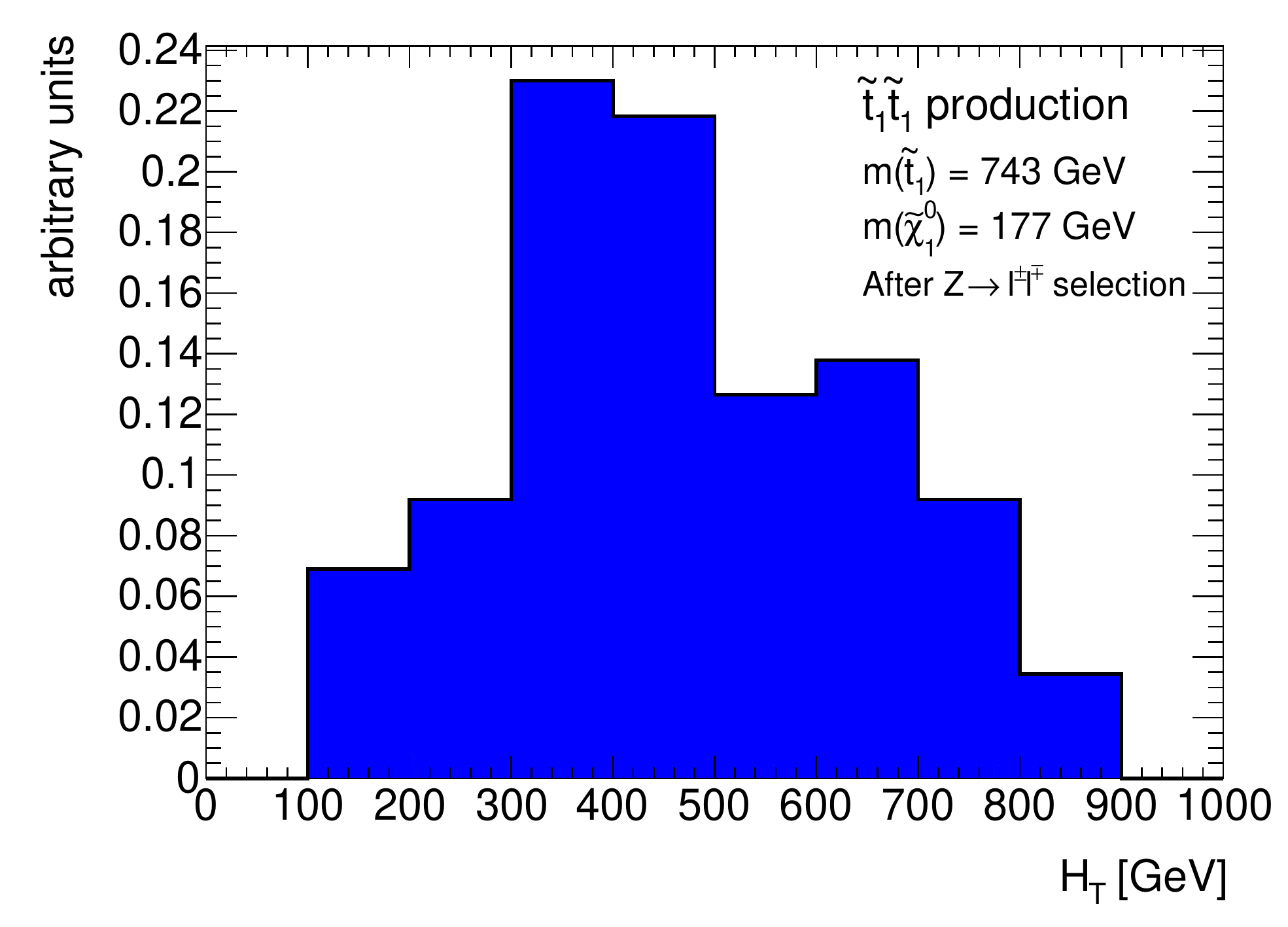}
  \caption{ $\HT$ distribution in arbitrary units from the decay of a pair of stops of $m_{\tilde t} =743$~GeV to jets plus \met after applying the selection $Z\to l^+ l^-$.}
\label{HTstop}
  \end{figure}
  \end{center}
Yet, we can also consider the strong production of stop pairs, where the current bounds on stop masses are much lower, $m_{\tilde t} \gtrsim 650$ GeV \cite{Khachatryan:2014doa,Aad:2014kra}. The total  $\tilde t \tilde t^*$ production cross-section is shown in Fig.~\ref{xstt}, calculated at NLL+NLO with NLL-fast. In this case, we can see that we could reach the required cross-section for $m_{\tilde t} \lesssim 750$ GeV which in principle could be achievable in general SUSY models (always assuming that every stop produces a $Z$-boson in its decay). However, the cuts $H_T > 600$~GeV and $E_T^{\rm miss} > 225$ are very restrictive. We can see this in Fig.~\ref{HTstop}, where we show the $\HT$ distribution from the decays of stop pairs with  $m_{\tilde t_1}= 750$~GeV. As can be seen here, the $\HT$ distribution peaks at $\HT \simeq$ 400--500~GeV as a consequence of the relatively small stop mass, and only a small fraction of the events are able to overcome the cut on $\HT$.  Therefore, we must conclude that it is not possible to generate the required cross-section and fulfil the requirements of the observed excess through stop production.  

\begin{center}
  \begin{figure}[h!]
  \includegraphics[scale=0.8]{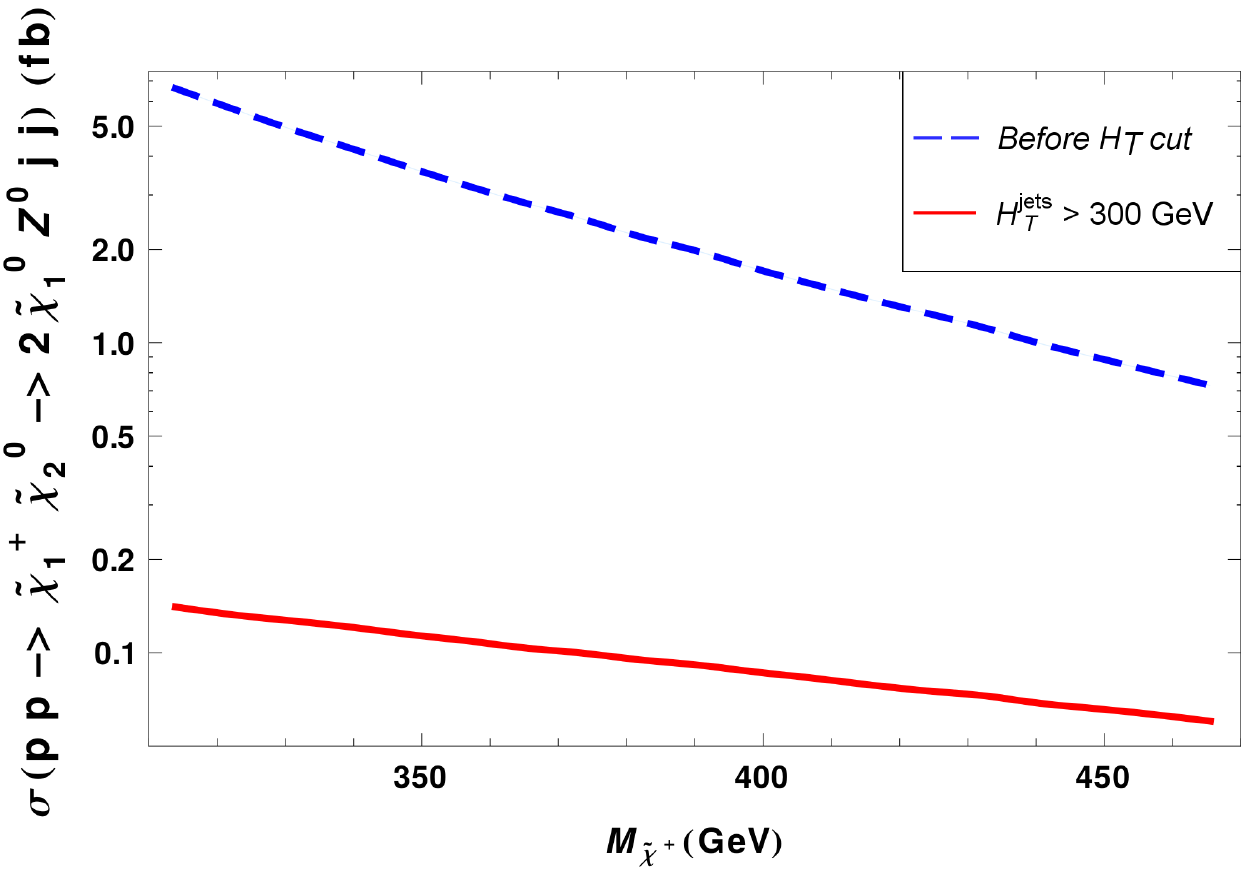}
  \caption{Production cross-section of a pair $\chi^0_2 \chi^\pm_1$ as a function of $m_{\tilde \chi^\pm_1}$ in the CMSSM with  $m_0=4$~TeV, $A_0=0$ and $\tan \beta=40$ and $\mu >0$. In blue (dashed), the cross-section with no cut on \met or $\HT$ and in red (solid), the cross-section after applying a cut on $H_T^{\rm jets} \geq 300$~GeV}
\label{xschichi0}
  \end{figure}
  \end{center}

Apart from the production cross-sections of coloured sparticles, we could still consider the weak production of charginos/neutralinos which can be large enough for light gauginos.  Taking into account that the current bounds on chargino and neutralino masses are not very stringent, electroweak production is worth exploring \cite{Khachatryan:2014mma,Aad:2014vma}. It is well-known that the largest electroweak production cross-sections are those corresponding to $\tilde W ^0 \tilde W^\pm$ and $\tilde W ^+ \tilde W^-$, in terms of gauge eigenstates. This would correspond to $\chi^0_2 \chi_1^\pm$ and  $\chi^+_1 \chi_1^-$ in terms of mass eigenstates. However, the observed signal requires $Z$-boson on-shell, at least two jets, $E^{\rm miss}_T > 225$~GeV and a minimum $H_T$ of 600~GeV. As an example, we present, in Figure \ref{xschichi0}, the production cross-section $\sigma ( p p \to \chi_1^+ \chi^0_2 \to (\chi^0_1~ W^+)~ (\chi^0_1~ Z) \to (\chi^0_1~j j)~(\chi^0_1~ Z^0) )$ in the case where the only restriction imposed is a minimum $p_T \geq 20$~GeV for the jets (in blue/dashed), and the same cross-section after applying a cut on the hadronic $H_{T}^{\rm jets} \geq 300$~GeV . As it can be seen, the cross-section with charginos of $m_{\chi^+_1} \lesssim 350$~GeV looks, in principle, able to accommodate the required $Z$ production. However, the situation changes after we impose the experimental cuts used in the analysis. The required $E_T^{\rm miss}$ is easily obtained if $m_{\chi^0_1} \gtrsim 150$~GeV, but the requirement on $H_T$ is very restrictive.
In fact, for chargino masses $m_{\chi^\pm_1} \simeq 350$~GeV, the largest contribution to the production cross-section would correspond to $\sqrt{\hat s} \simeq 700$~GeV, which can never produce $H_T$ of 600~GeV and  $E^{\rm miss}_T > 225$~GeV. Then, we can expect the cross-section  $\sigma(p p \to \chi_1^0 \chi^+)$ at larger $\sqrt{\hat s}$ values to be  strongly reduced. This is shown by the red (solid) line in  Figure \ref{xschichi0}: a relatively mild cut on the hadronic $\HT$ reduces the cross-section by more than one-order of magnitude.    
Thus we must conclude that, although electroweak production could contribute efficiently to the production of additional $Z$-bosons, these events can not overcome the experimental cuts and can not give rise to the observed signal.    

With this channel, we have reviewed all relevant production cross-sections of different supersymmetric particles that could potentially explain the signal. The next step would be to calculate the average number of $Z$-bosons per parent particle $Y$, that we use in Eq.~(\ref{requiredXS}).  

\subsection{Decay of Supersymmetric particles to $Z$-bosons}

$Z$ bosons are produced through the decay chains of most MSSM particles, although the number of $Z$-bosons obtained per each supersymmetric particle produced depends on the identity of the supersymmetric particle initially produced and on the supersymmetric spectrum below its mass. The main sources of $Z$-bosons are the decays of neutralinos and charginos and also in some squark decays. The couplings of charginos/neutralinos to $Z$ are given by,
  \begin{eqnarray}
{\cal L}_{Z \chi\chi} &=& \frac{g_2}{\cos \theta_W} ~Z_\mu \left[ \chi^+_i  \gamma^\mu \left( O_{ij}^{\prime L} P_L + O_{ij}^{\prime R} P_R\right) \chi^-_j + \bar\chi^0_i  \gamma^\mu \left( O_{ij}^{\prime \prime L} P_L + O_{ij}^{\prime \prime R} P_R\right) \chi^0_j \right] \\
\mbox{ with}&~&O_{ij}^{\prime L} = - V_{i1} V^*_{j1} - \frac{1}{2}  V_{i2} V^*_{j2} + \delta_{ij} \sin^2 \theta_W \nonumber \\
&~&O_{ij}^{\prime R} = - U^*_{i1} U_{j1} - \frac{1}{2}  U^*_{i2} U_{j2} + \delta_{ij} \sin^2 \theta_W \nonumber \\
&~&O_{ij}^{\prime \prime L} = - \frac{1}{2}  N_{i3} N^*_{j3} + \frac{1}{2}  N_{i4} N^*_{j4} \nonumber \\
&~&O_{ij}^{\prime \prime R} = - O_{ij}^{\prime \prime L} \nonumber \, .
  \end{eqnarray}
Therefore, we can obtain $Z$ bosons in the decays of higgsino-like neutralinos and charginos. For instance in a usual mSugra spectrum, the second neutralino will only produce $Z$-bosons through its (relatively small) higgsino component while the two heavier neutralinos can be expected to produce a sizeable number of $Z$-bosons. On the other hand, charginos can produce $Z$-bosons both through the wino and from the higgsino component but only in decays of the heavier charginos, as the lightest one will only decay to a $W$-boson and a neutralino (or lepton-slepton if $m_{\tilde l}\leq m_{\chi^+}$).

\begin{center}
  \begin{figure}[h!]
  \includegraphics[scale=0.6]{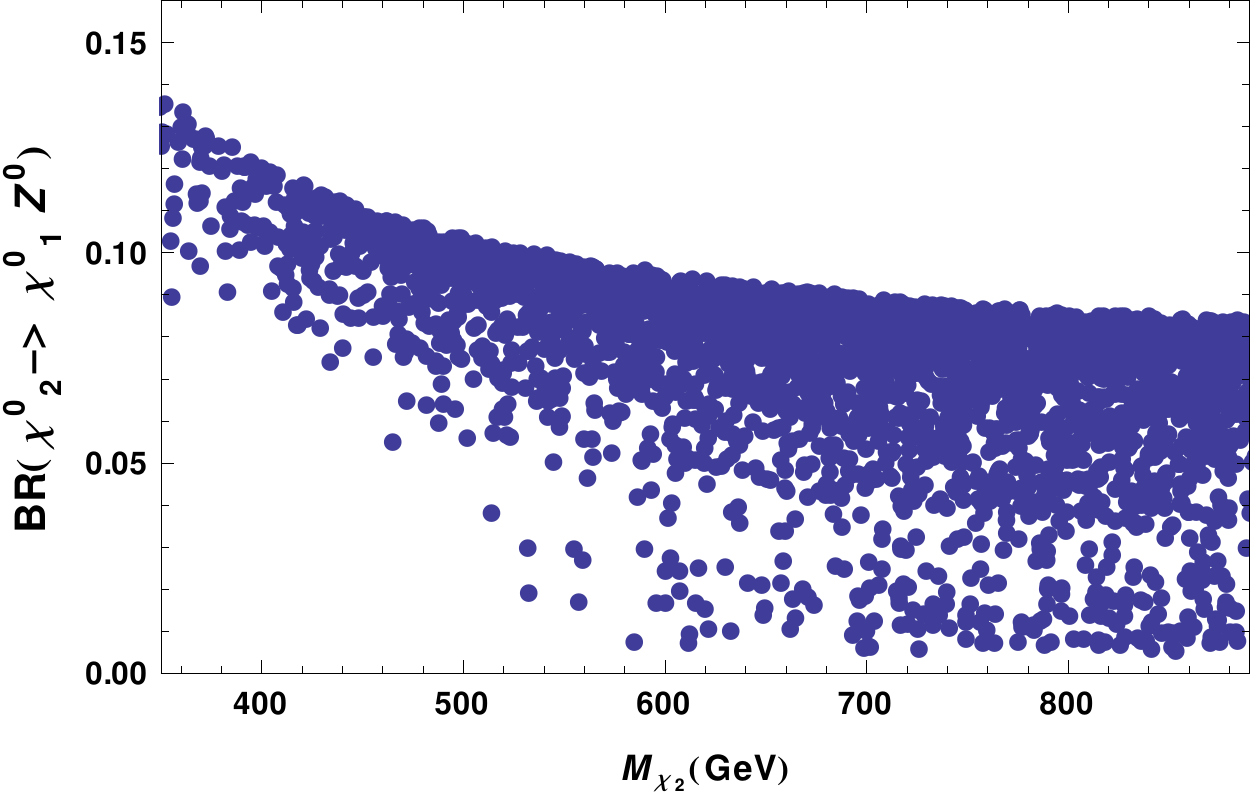} \includegraphics[scale=0.6]{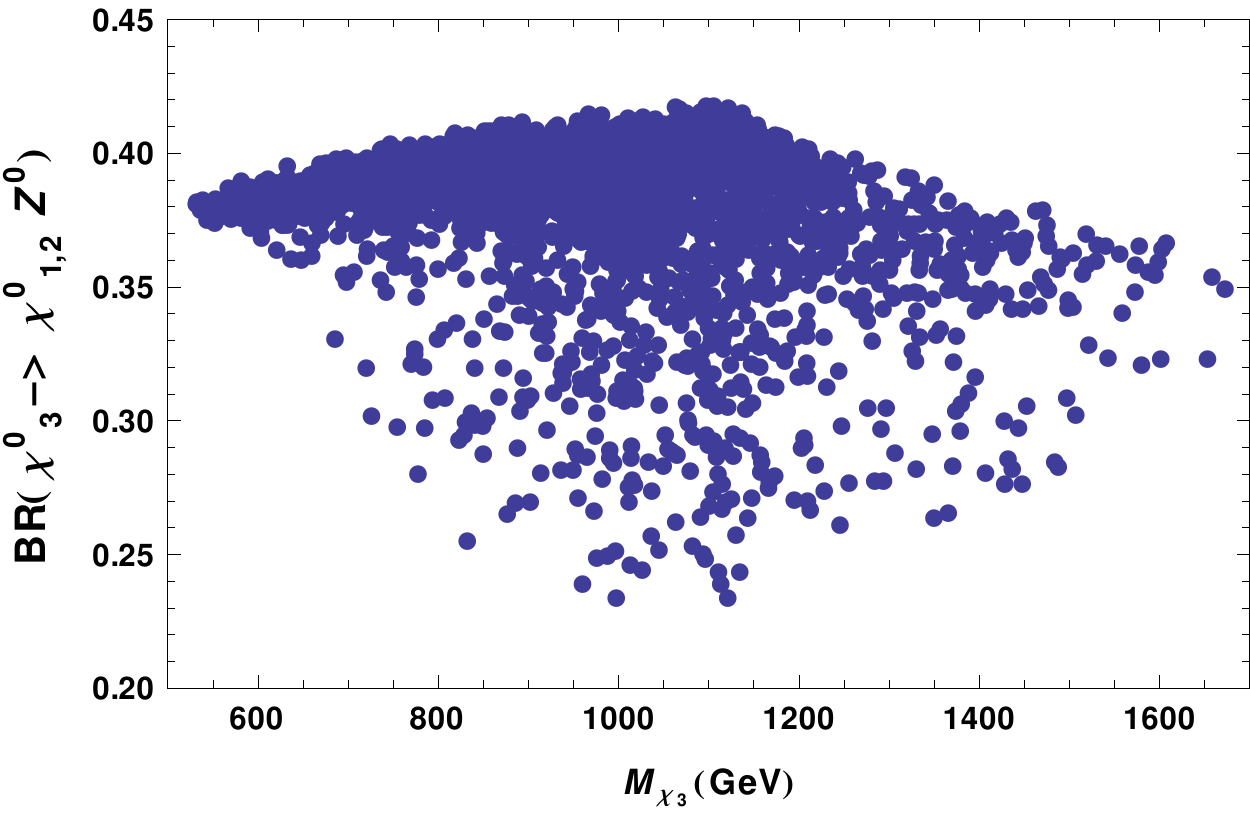}
  \caption{Average number of $Z$ bosons, ${\cal N}(Y\to Z)$, in the decays of $\chi^0_2$ (left) and $\chi^0_3$ (right) for a typical mSugra spectrum.}
\label{BRchi0}
  \end{figure}
  \end{center}

\begin{center}
  \begin{figure}[h!]
  \includegraphics[scale=0.6]{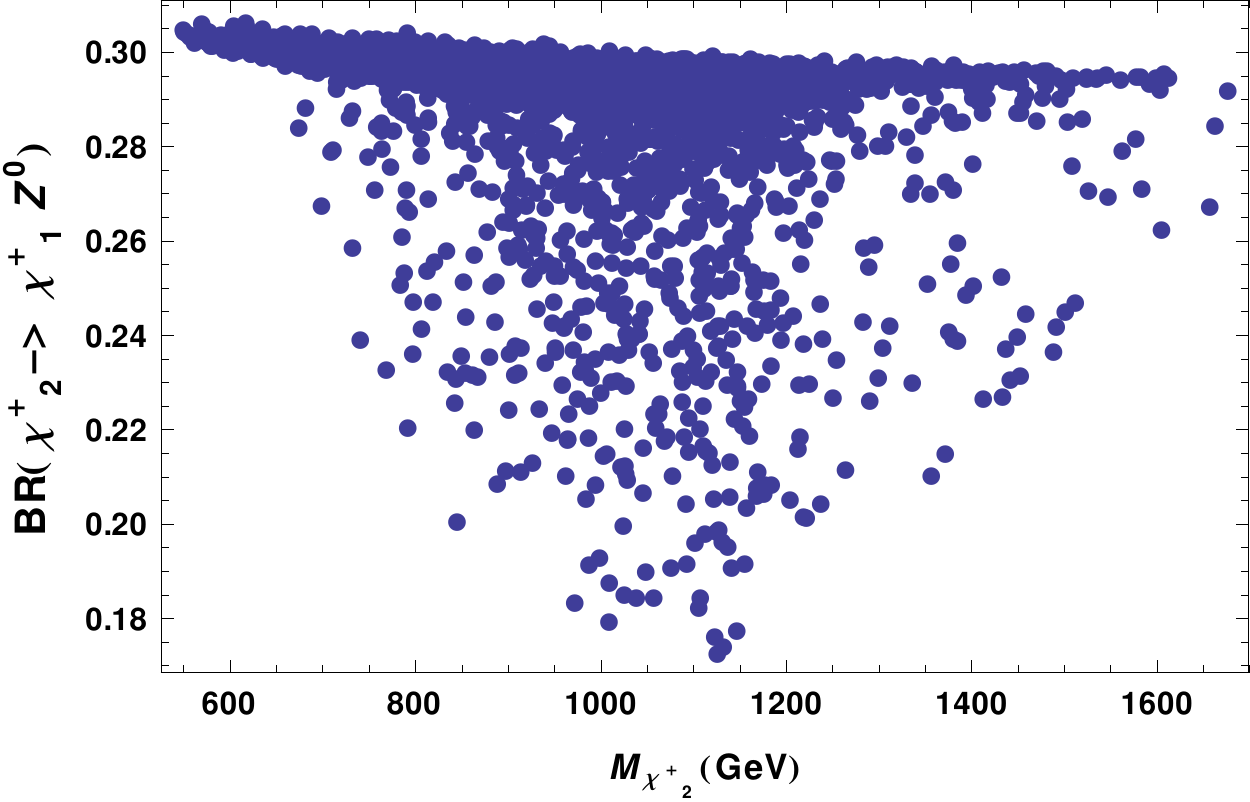}
  \caption{Average number of $Z$ bosons, ${\cal N}(\chi^+_2 \to Z)$, in the decays of $\chi^+_2$  for a typical mSugra spectrum}
\label{BRchip}
  \end{figure}
  \end{center}

In Fig.~\ref{BRchi0} we can see the values of ${\cal N}(\chi^0_{2,3}\to Z )$ as a function of $m_{\chi^0_{2,3}}$ in an mSugra model. Here, we obtain ${\cal N}(\chi_2 \to Z)$ around 0.1, as expected if the higgsino content is relatively small. Then, ${\cal N}(\chi_3 \to Z)$ can reach at most 0.45 while the other 50\% of the decays go to $\chi^\pm W^\mp$. Similarly the heavy charginos produce $Z$-bosons in their decay as can be seen in Fig.~\ref{BRchip}. In this case ${\cal N}(\chi^+_2 \to Z)$ can be 0.3, while we have similar branching ratios to $\chi_1^0 W^+$ and $\chi_1^+ h$. 
Besides chargino and neutralino decays, $Z$-bosons couple also to sfermions through,
\begin{eqnarray}
{\cal L}_{Z \tilde q} &=& \frac{-i g_2}{\cos \theta_W}(T_{3 I} - e_I \sin^2 \theta_W) ~Z_\mu \,\tilde q_\alpha^* \overleftrightarrow{\partial}^\mu  \tilde q_\beta ~ \left( (\Gamma_{q L}^{\rm SCKM})_{I \alpha}^* (\Gamma_{q L}^{\rm SCKM})_{I \beta} +(\Gamma_{q R}^{\rm SCKM})_{I \alpha}^* (\Gamma_{q R}^{\rm SCKM})_{I \beta} \right) \, . \nonumber \\
\end{eqnarray}
As we can see, the $Z$ couplings are chirality diagonal and therefore, in decays, they can only enter through chirality mixing. Although these couplings could be flavour changing, flavour mixing is bounded to be small due to the stringent flavour changing neutral current constraints. Therefore we can expect a sizeable amount of $Z$-bosons produced only through chirality mixing in third generation sfermions in decays like $\tilde t_2 \to \tilde t_1 + Z$, or $\tilde b_2 \to \tilde b_1 + Z$ in the large $\tan \beta$ regime. 

In addition, we obtain $Z$-bosons in the decay chains of strongly produced sparticles. We can obtain $Z$-bosons at different steps of the decay chain, either through the couplings of $Z$ to sfermions or to charginos/neutralinos that we saw above. For instance, if we produce a pair of gluinos a possible decay chain 
would be $\tilde g \to \tilde t_2 t \to \tilde t_1 Z~ t \to \chi_2^+ b~ Z~ t \to
\chi_1^+ Z~ b~ Z~ t \to \chi_1^0 W^+~Z~ b~ Z~ t$. Therefore taking
into account the corresponding branching ratios, this decay chain would 
contribute with $2 \times \mbox{BR}(\tilde g \to \chi_1^0 W^+~Z~ b~ Z~ t)$ to
${\cal N}(\tilde g \to Z )$, the number of $Z$-bosons produced per $\tilde g$
  produced.
 \begin{center}
  \begin{figure}[h!]
  \includegraphics[scale=0.62]{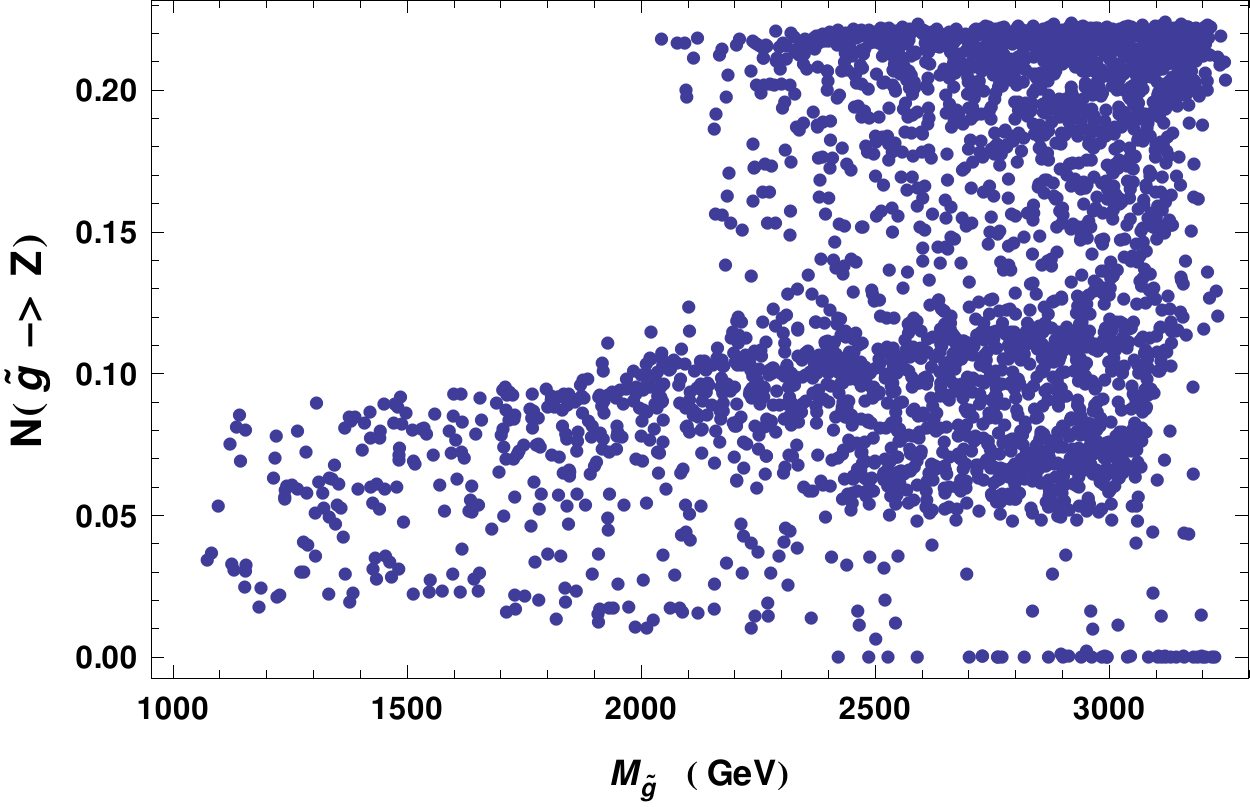} \includegraphics[scale=0.62]{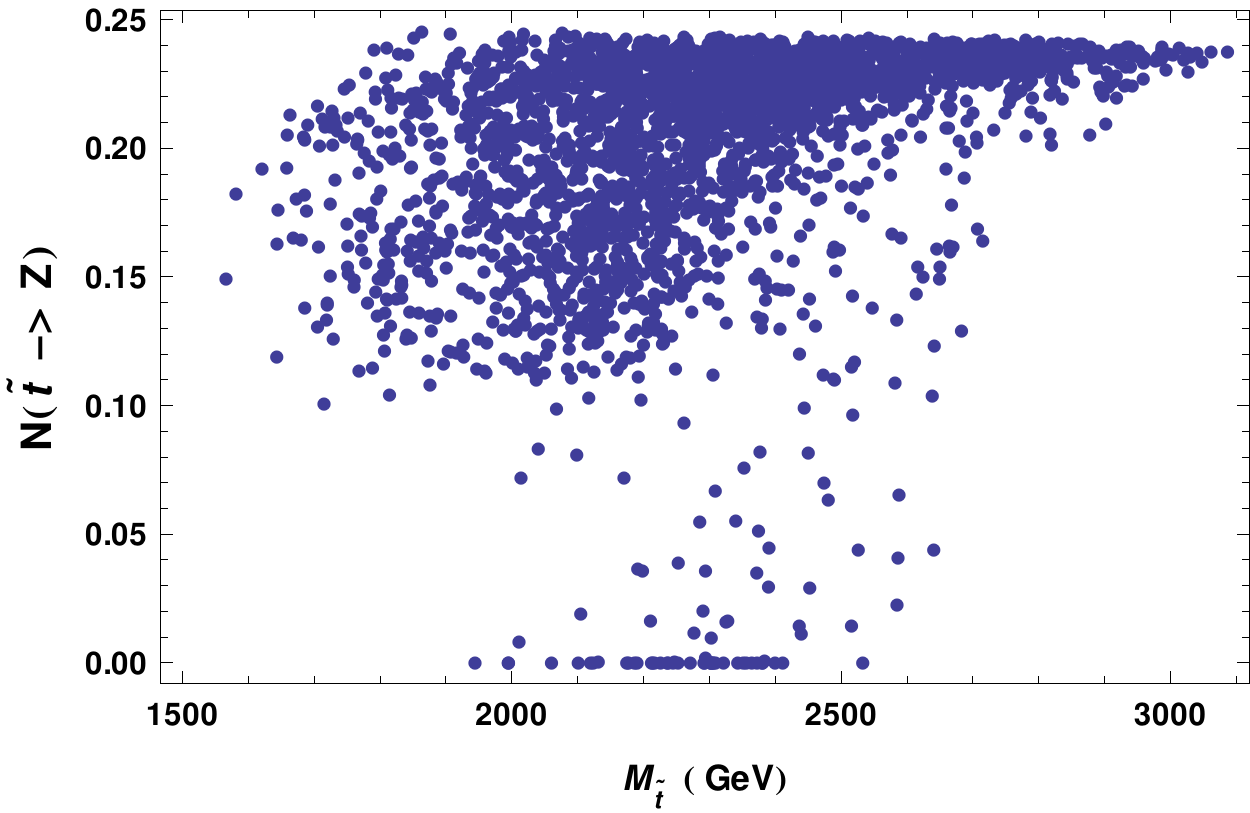}
  \caption{Average number of $Z$ bosons, ${\cal N}(Y \to Z)$, in gluino (left) and stop (right) decays for a typical mSugra spectrum.}
  \label{gluino-stop}
  \end{figure}
  \end{center}
As we can see in Fig.~\ref{gluino-stop}, in a typical MSSM spectrum we obtain at most 0.2 $Z$-bosons per stop or gluino while other squarks produce far fewer $Z$-bosons per squark. 
Although Figs. \ref{BRchi0}--\ref{gluino-stop} have been obtained from a mSugra spectrum, the expected number of $Z$ bosons would be very similar in other MSSM versions, as it depends only on the spectrum below the mass of the originally produced $Y$ particle and the content of charginos-neutralinos. 

As shown in the previous section, the only possibility to explain the signal is to produce a stop or a gluino-pair as the lightest coloured sparticle, being all other squarks much heavier and only neutralinos, charginos and possibly some sleptons can be below the gluino or stop mass. Moreover, given the size of production cross-sections consistent with the present searches, we need to obtain nearly one $Z$-boson per $Y$ particle produced. Therefore, from the expected numbers of $Z$-bosons that we have seen in this section, we have to conclude that it is not possible to reproduce the observed signal in a MSSM with a stable (and light) neutralino. 
 
Nevertheless, we can still consider different variations of the MSSM: 
\begin{itemize}
\item 
A first possibility would be to have a light gluino below 1 TeV that can evade the bounds from jets plus missing $E_T$ if it decays to a sufficiently heavy neutralino LSP in a sort of compressed spectrum. Under these conditions the gluino would be abundantly produced at LHC and even a small number of $Z$-bosons per gluino could fulfil the requirements to explain the observed signal. However, this would require strongly non-universal gaugino masses and very heavy LSP's and we will not follow this path here.
\item A second option is to consider an MSSM where the lightest neutralino decays to a lighter gravitino plus some $Z$-boson. This is the case in gauge mediated MSSM \cite{Dine:1993yw,Dine:1994vc,Dine:1995ag,Giudice:1998bp} and it could be also possible in gravity mediated MSSM if the gravitino is lighter than the neutralino which then becomes the NLSP \cite{Feng:2004mt}. In this case, the neutralino decays to $Z$-boson and gravitino if it is allowed by phase-space and the branching ratio will depend on the lightest neutralino composition. This is the possibility we will explore in the following.
\end{itemize} 
Thus, we will analyze a situation where the neutralino is the next to lightest supersymmetric particle and the LSP is the gravitino. All supersymmetric particles will decay to the lightest neutralino which then decays to gravitino plus a photon, a $Z$-boson or a Higgs. The decay width of the lightest neutralino to photon, $h$ or $Z$ plus gravitino \cite{Moroi:1995fs,Ellis:2003dn,Feng:2004mt} is given by,
\begin{eqnarray}
\Gamma(\chi_1^0 \to \gamma \tilde G) &=& \frac{\left|N_{11} \cos \theta_W +N_{12} \sin \theta_W\right|^2}{48\pi M_{Pl}^2} \frac{m_\chi^5}{m_{\tilde G}^2} \left[1 -\frac{m_{\tilde G}^2}{m_\chi^2}\right] \left[1 +3\frac{m_{\tilde G}^2}{m_\chi^2}\right] \, , \nonumber \\
\Gamma(\chi_1^0 \to Z \tilde G) &=& \frac{\left|-N_{11} \sin \theta_W +N_{12} \cos \theta_W\right|^2}{48\pi M_{Pl}^2} \frac{m_\chi^5}{m_{\tilde G}^2} F(m_\chi,m_{\tilde G},m_Z)\, , \nonumber \\
\Gamma(\chi_1^0 \to h \tilde G) &=& \frac{\left|-N_{13} \sin \alpha +N_{14} \cos \alpha\right|^2}{96\pi M_{Pl}^2} \frac{m_\chi^5}{m_{\tilde G}^2} F(m_\chi,m_{\tilde G},m_h)\, ,
\end{eqnarray}
with $F(x,y,z)$ a function of the particle masses non relevant for our discussion that can be obtained from \cite{Feng:2004mt}.
As we can see, if the lightest neutralino is bino-like, $N_{11}\simeq 1$, and the mass difference between neutralino and gravitino is larger that the $Z$-mass, the branching ratios are BR$(\chi_1 \to \tilde G \gamma )\simeq \cos^2 \theta_W \simeq 0.8$ and  BR$(\chi_1 \to \tilde G $Z$ )\simeq \sin^2 \theta_W \simeq 0.2$. Similarly if the lightest neutralino is wino-like, the branching ratios get exchanged. From this equation we can also see that it is possible to get a very large BR to $Z$ bosons as needed to reproduce the observed signal if $(-N_{11} \sin \theta_W +N_{12} \cos \theta_W) \simeq 1$, but this is only possible if the lightest neutralino has a very large wino component.

Although in gauge mediation models the LSP is always the gravitino, the gaugino masses in minimal models are proportional to the gauge couplings and therefore the LSP is mostly bino with small wino and higgsino components. Then, we will have to consider other extensions of the gauge mediation idea, like the so-called General Gauge Mediation (GGM) where the gaugino and sfermion masses depend on hidden sector current correlators which can be different for different gauge groups or particle representations. As we will show in the following section, in these models it is possible to obtain general neutralino NLSP as required to explain the excess.

\section{A possible explanation in General Gauge Mediation} 
\label{GGM}
Minimal gauge mediation predicts that all scalar and gaugino masses originate from a single scale and powers of the gauge couplings \cite{Giudice:1998bp}. Recently
a model-independent generalization of gauge mediation was proposed under the name of General Gauge Mediation \cite{Meade:2008wd,Buican:2008ws}, where all the dependence of soft masses on the hidden sector is encoded in three real and three complex parameters obtained from a small set of current-current correlators. In these models the gaugino and sfermion masses are given by,
 \begin{eqnarray}
M_r &=& g_r^2 M_s \tilde B_r^{1/2}(0) \\
m^2_{\tilde f} &=& g_1^2 Y_f \zeta + \sum_{r=1}^3 g_r^2 {\cal C}_2(f|r) M_s^2 \tilde A_r \nonumber \,,
  \end{eqnarray}
with
\begin{equation}
\tilde A_r = -\frac{1}{16 \pi^2} \int dy \left( 3 \tilde C_1^{(r)}(y) -4 \tilde C_{1/2}^{(r)}(y)+ \tilde C_0^{(r)}(y) \right)\,,
\end{equation}
$\tilde B_r^{1/2}(0)$, $ C_\rho^{(r)}(y)$ (with $\rho=0,1/2,1$, corresponding to scalar, fermion and vector) are associated with the current-current correlators in the hidden sector, $\zeta$ is a possible Fayet-Illiopoulos term ($\zeta=0$ in the following), ${\cal C}_2(f|r)$ the quadratic Casimirs and $M_s$ a characteristic SUSY-breaking scale in the hidden sector.

Having six parameters, $ \tilde B_r^{1/2}(0)$ and $\tilde A_r$, to fix the soft masses in the observable sector, it is clear now that we have much more freedom in GGM \cite{Carpenter:2008he,Rajaraman:2009ga,Thalapillil:2010ek,Kats:2011qh} and, in particular, we have
\begin{equation}
\frac{M_1}{g_1^2} \neq \frac{M_2}{g_2^2} \neq  \frac{M_3}{g_3^2} \,,
\end{equation}
as required to reproduce the observed signal at ATLAS. In particular, we need the NLSP to decay to gravitino and a $Z$-boson with a branching ratio close to one. Fortunately, this is possible in GGM as shown in Ref.~\cite{Ruderman:2011vv,Kats:2011qh}.

In this GGM scenario we have used SPheno-3.3.3 \cite{Porod:2003um,Porod:2011nf} to obtain the full supersymmetric spectrum at LHC energies. We define  a first parameter point GGM1 with the following parameters, $M_s = 400$~TeV, $\tilde B_1^{1/2}=\tilde A_1=309$~TeV, $\tilde B_2^{1/2}=\tilde A_2 =151$~TeV, $\tilde B_3^{1/2}=129$~TeV, $\tilde A_3=316$~TeV and $\tan\beta=9.8$. With these parameters we obtain the spectrum shown in Table \ref{tab1}.
\begin{table}
\begin{center}
\begin{tabular*}{1.00\textwidth}{@{\extracolsep{\fill}}|c c c c c c c c c|}
\hline \hline
{\rm Particle}& $\tilde g$& $\chi_1^0$ & $\chi_2^0$ & $\chi_3^0$  & $\chi_4^0$ & $\chi_1^\pm$ & $\chi_2^\pm$& $\tilde G$~~~ \\
\hline
{\rm Mass (GeV)} & 1088.0 & 428.4 & 431.34 & 1357.0 & 1360.9 & 429.1 & 1361.0 & $4.8 \times 10^{-9}$~~~ \\
\hline \hline
{\rm Particle}&  $\tilde q_L$ & $\tilde q_R$ & $\tilde b_1$ & $\tilde b_2$ & $\tilde t_1$ &  $\tilde t_2$ & $\tilde l_L$& $\tilde l_R$~~~ \\
\hline
{\rm Mass (GeV)} & 3006 & 2957 & 2876 & 2952 & 2716 & 2881 & 5863 & 5328~~~ \\
\hline\hline
{\rm Particle}&  $h$ & $H$ & $A$ & $H^+$ & & & & ~~~ \\
\hline
{\rm Mass (GeV)} & 119.4 & 1471 & 1471 & 1473 & & & & ~~~ \\
\hline
\end{tabular*}
\end{center}
\caption{\label{tab1} SUSY spectrum in the GGM1 parameter point.}
\end{table}
With respect to this spectrum, some comments are in order: 
\begin{enumerate} 
\item The two lightest neutralinos and the lightest chargino are very similar in mass, $\sim 430$~GeV and this allows a large neutrino mixing as required.  In fact the neutralino mixing matrix is given by
\begin{eqnarray}  
\label{Nmix}
N_{ij}\simeq\left( 
\begin{array}{cccc}
-0.51 & 0.85& -0.076& 0.031 \\ 
0.86 & 0.51& -0.0024& 0.0071 \\ 
-0.015 & 0.028& 0.71& 0.71\\
-0.037 & 0.065& 0.70& -0.71
\end{array}%
\right)
\end{eqnarray}
On the other hand the relatively large NLSP mass is needed to overcome the \met cut.
\item The gluino is relatively light $m_{\tilde g} = 1088.0$~GeV which allows
 for a sizeable production cross-section and taking into account the squark masses of order $\sim 3$~TeV.  As we can see  in Ref. \cite{Aad:2014wea}, this mass would be allowed at 1 $\sigma$ in a simplified MSSM with gluino-squark-netralino.
\item The lightest Higgs mass must reproduce the observed value at LHC of $m_h\simeq 125$~GeV and in this spectrum it reaches only 119.4 GeV. This problem (typical in minimal gauge mediation models) can be solved either by increasing the stop masses taking a larger $\tilde A_3$ or assuming extra operators in the Higgs sector, as the dimension 5 operators proposed by Dine, Seiberg and Thomas \cite{Dine:2007xi}. Here, we assume that this problem is solved by one of these mechanisms, given it does not affect the observed phenomenology on the $Z$-peak 
\end{enumerate}
 \begin{center}
  \begin{figure}[h!]
  \includegraphics[scale=0.4]{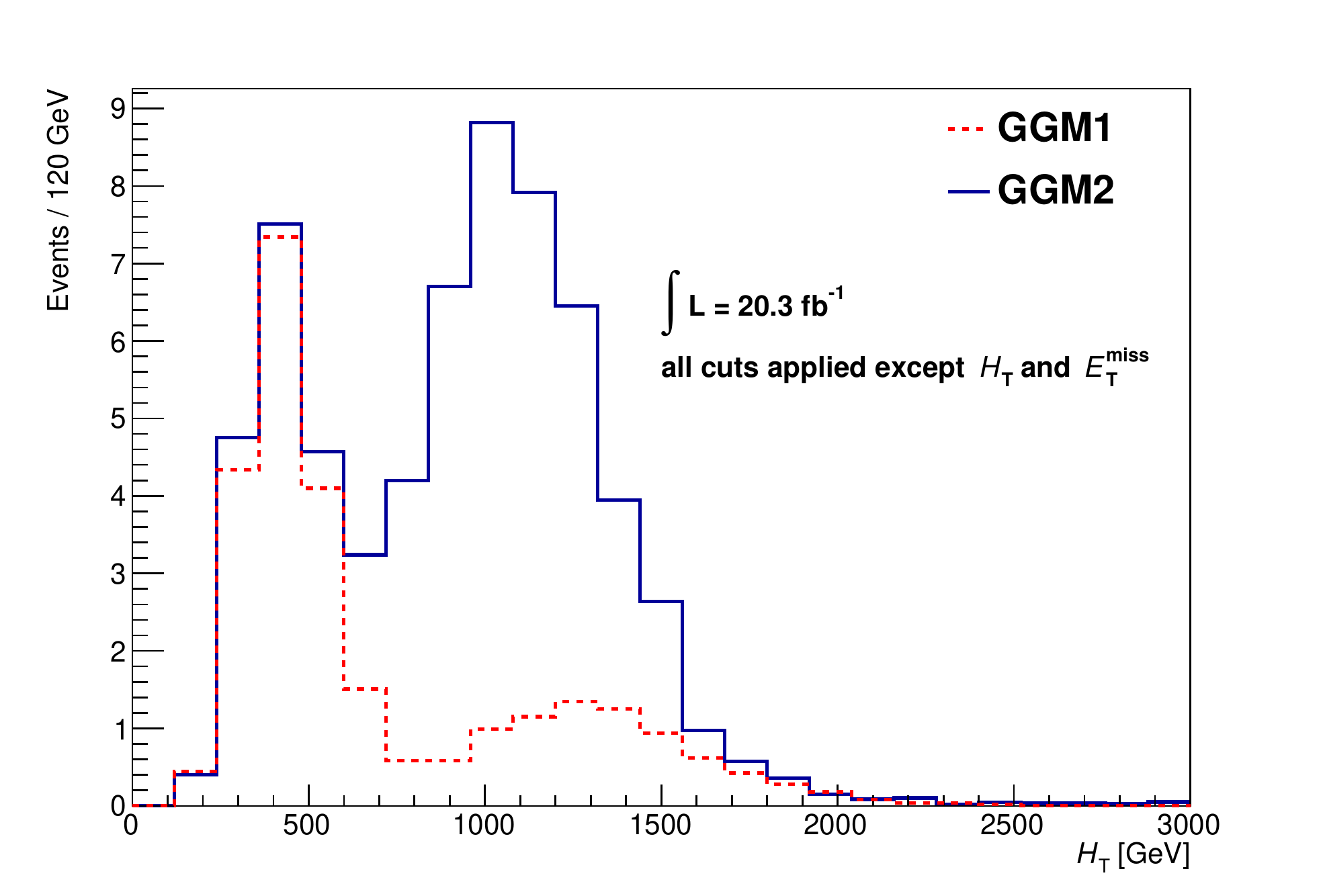} \includegraphics[scale=0.4]{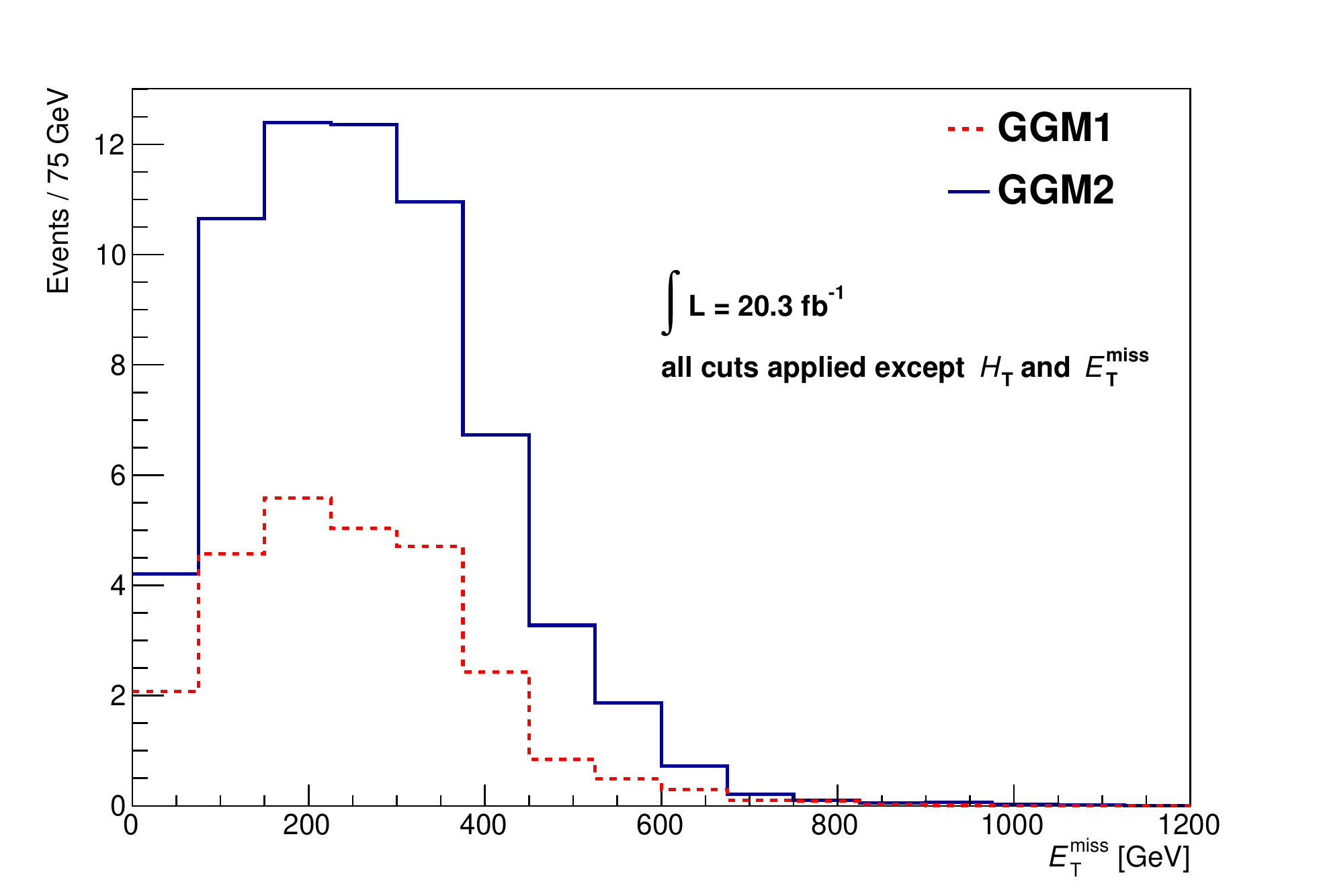}
  \caption{$\HT$ (left) and \met (right) distributions after applying the selection detailed in section~\ref{ATLAS_selection}, except for the cuts on \HT and \met . The GGM1 point is represented by the dashed red line and the GGM2 point, by the solid blue line.}
  \label{GGMHTMET}
  \end{figure}
  \end{center}

\begin{center}
  \begin{figure}[h!]
  \includegraphics[scale=0.5]{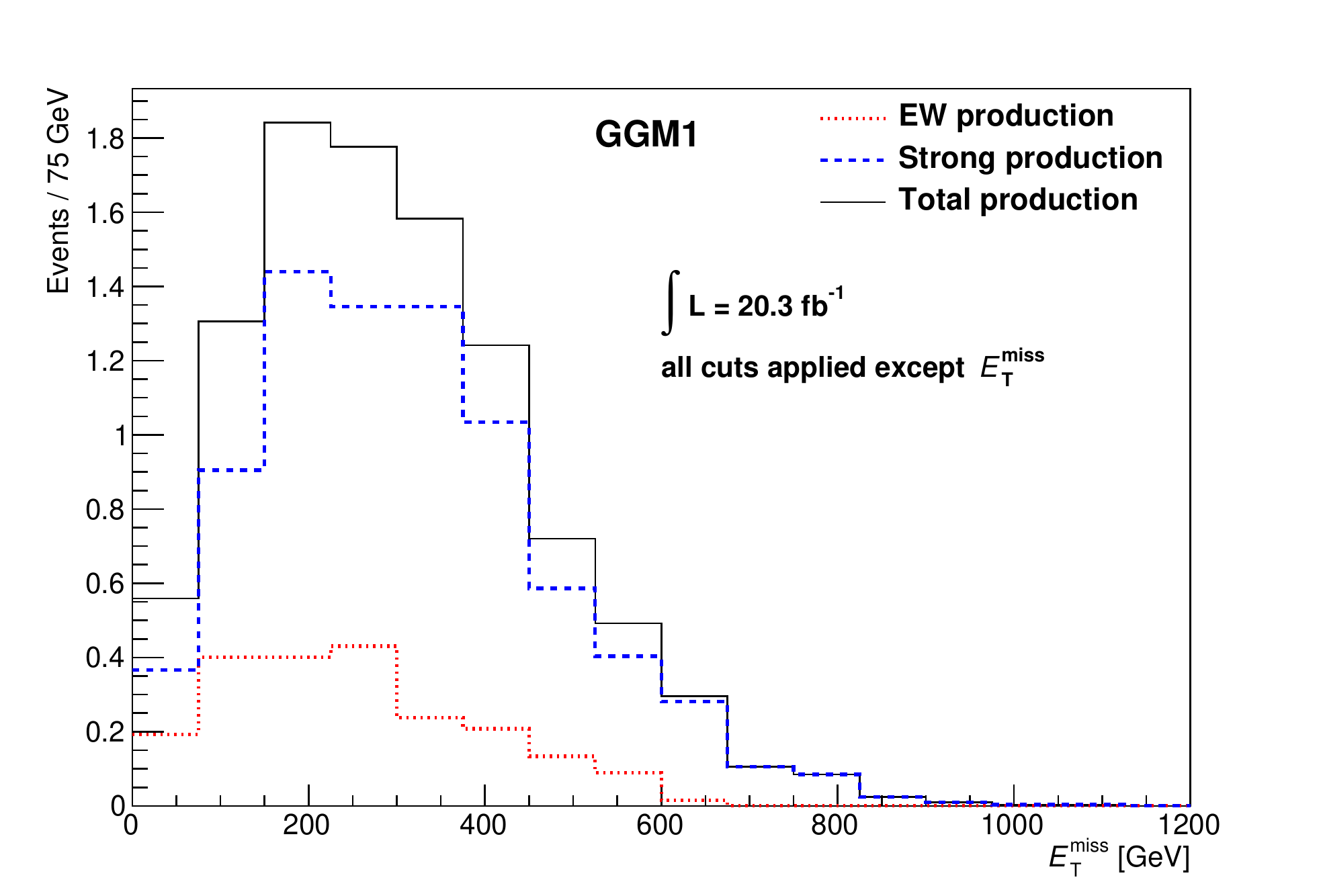} 
  \caption{\met distribution corresponding to the GGM1 point from strong production (solid blue line) and electroweak production (dashed red line) with the same selection as in Fig.~\ref{GGMHTMET}, after applying the $\HT>600~\GeV$ cut.}
  \label{GGM1}
  \end{figure}
  \end{center}
Under these conditions, the lightest neutralino width is $\Gamma_{\chi_1} = 4.097 \times 10^{-10}$~GeV and the decay branching ratios are BR$(\chi^0_1 \to \tilde G \gamma)=1.14 \times 10^{-3}$,  BR$(\chi^0_1 \to \tilde G Z)= 0.997$ and  BR$(\chi^0_1 \to \tilde G h)=1.35 \times 10^{-3}$. Therefore, gluinos are produced at LHC with a cross-section of $8.4 \pm 1.6$~fb at NLL+NLO and after going through different decay chains all of them produce a $Z$-boson plus a gravitino. In this case, the strong production represents approximately $20\%$ of the total production cross-section.

We simulate the production of supersymmetric particles at LHC at 8 TeV (LHC8) with this spectrum using Pythia 8.1 \cite{0710.3820} with Prospino2 \cite{Beenakker:1996ch,Beenakker:1997ut,Beenakker:1999xh} K-factors and the response of the ATLAS detector using Delphes \cite{deFavereau:2013fsa}. 
The selection of events for this study is done as close as possible to that performed in ATLAS~\cite{Aad:2015wqa}. The dashed red line in Fig.~\ref{GGMHTMET} and Fig.~\ref{GGM1} shows the \HT and \met distributions respectively for the GGM1 point after applying all the selection detailed in section~\ref{ATLAS_selection} except for the \HT and \met cuts. 
In the $\HT$ distribution, we can distinguish the two peaks corresponding to electroweak production at low \HT values and gluino production at higher \HT . From here, we can expect that the cut on $\HT$ will eliminate most of the electroweak production but not the gluino production.
This can be seen in Fig.~\ref{GGM1}, where the \met distribution is presented separately for strongly produced events (solid blue line) and for the electroweak component of the production (dashed red line) for the same selection as in Fig.~\ref{GGMHTMET} but after applying the $\HT$ cut, i.e. the final selection except for the cut on \met . The electroweak component is significantly reduced by the \HT cut while mainly only events coming from strong production survive the cut, as expected.  In fact, in this simulation of point GGM1, 99\% of the gluino points and only 11\% of the electroweak points have survived the \HT cut.    
We see that the peak in the \met distribution is approximately at $m_{\chi_1^0}/2$, and, for $m_{\chi_1^0}= 425$~GeV, a reasonable fraction of the events will survive the \met cut at 225 GeV. In the simulation, 65\% of the gluino point and 53\% of the electroweak points survive this cut. However, due the relatively small production cross-section, the final number of events is small. 
In this simulation and after applying all relevant experimental cuts, an expected signal of $6.34 \pm 1.02$ lepton pairs is found, to be compared with the observed excess of  $19.4 \pm 3.2$. This number of surviving events was obtained at NLO with Pythia and Prospino2 but, unfortunately, it is still too low to explain the observations.

Trying to obtain a model able to account for the excess, we consider a second point in our GGM scenario with a lighter gluino.
The GGM2 point is obtained with $M_s = 400$~TeV, $\tilde B_1^{1/2}=\tilde A_1=309$~TeV, $\tilde B_2^{1/2}=\tilde A_2 =150$~TeV, $\tilde B_3^{1/2}=110$~TeV, $\tilde A_3=270$~TeV and $\tan\beta=9.8$. With these parameters we obtain the spectrum shown in Table \ref{tab2}.
\begin{table}
\begin{center}
\begin{tabular*}{1.00\textwidth}{@{\extracolsep{\fill}}|c c c c c c c c c|}
\hline \hline
{\rm Particle}& $\tilde g$& $\chi_1^0$ & $\chi_2^0$ & $\chi_3^0$  & $\chi_4^0$ & $\chi_1^\pm$ & $\chi_2^\pm$& $\tilde G$~~~ \\
\hline
{\rm Mass (GeV)} & 911.4 & 424.9 & 432.7 & 1111.8 & 1117.1 & 425.8 & 1117.2 & $4.8 \times 10^{-9}$~~~ \\
\hline \hline
{\rm Particle}&  $\tilde q_L$ & $\tilde q_R$ & $\tilde b_1$ & $\tilde b_2$ & $\tilde t_1$ &  $\tilde t_2$ & $\tilde l_L$& $\tilde l_R$~~~ \\
\hline
{\rm Mass (GeV)} & 2510 & 2470 & 2400 & 2450 & 2250 & 2400 & 5890 & 5360~~~ \\
\hline\hline
{\rm Particle}&  $h$ & $H$ & $A$ & $H^+$ & & & & ~~~ \\
\hline
{\rm Mass (GeV)} & 118.1 & 1250 & 1250 & 1253 & & & & ~~~ \\
\hline
\end{tabular*}
\end{center}
\caption{\label{tab2} SUSY spectrum for the point GGM2.}
\end{table}
The neutralino mixing matrix in point GGM2 is similar to the corresponding mixing matrix in GGM1, and the BR($\chi^0_1 \to \tilde G Z$) = 0.94. However, gluino is now much lighter and the gluino-gluino cross-section is now $(41.6 \pm 7.5)$ with Prospino at NLL+NLO, thus we can expect many more gluino pairs to be produced and a larger contribution in the final selection for this point.

\begin{center}
  \begin{figure}[h!]
  \includegraphics[scale=0.5]{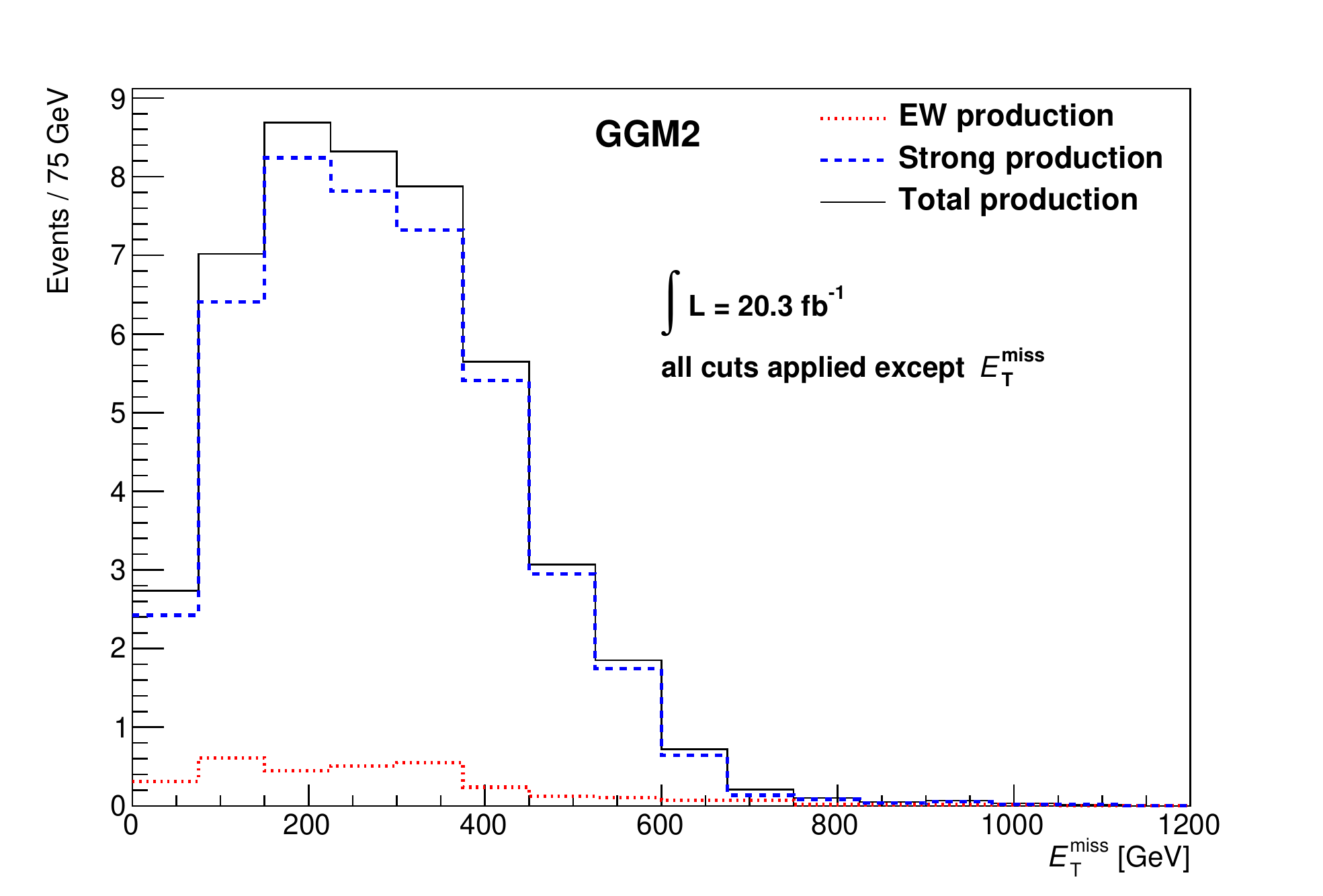}
  \caption{ \met distribution corresponding to the GGM2 point from strong production (solid blue line) and electroweak production (dashed red line) with the same selection as in Fig.~\ref{GGMHTMET}, after applying the $\HT>600~\GeV$ cut.}
  \label{GGM2}
  \end{figure}
  \end{center}

The simulation for this GGM2 point is presented by the solid blue line in Fig.~\ref{GGMHTMET}. The $\HT$ distribution peaks at slightly lower values than in the case of GGM1, due to the slightly lower gluino mass, but it is still enough to overcome the $\HT$ cut at 600 GeV. The \met distributions are similar for both GGM points, due to the very similar neutralino masses in both cases. However, in the case of GGM2 the strong production cross-section is larger and much more important in relation with the electroweak production: for the GGM2 point, the strong production represents $\sim 55\%$ of the total cross-section. As seen in Fig.~\ref{GGM2}, after applying the selection we obtain an expected signal of $28.0 \pm 4.7$ events,  slightly over the excess reported by ATLAS, showing that a signal point defined along these characteristics can be able to reproduce the excess.

We have to emphasize here that it is not difficult to obtain the observed excess for light gluino masses, and a gluino mass in between the two presented examples, $m_{\tilde g} \in [900,1100]$ GeV, could reproduce the observed signal.  However, these points may be in conflict with direct searches of jets plus \met \cite{Khachatryan:2015pwa,Aad:2015mia,Aad:2014wea}. There is a tension between this excess and the bounds from gluino searches in jets plus \met.
To quantify more accurately this tension, we use the program Checkmate \cite{Drees:2013wra} that allows to compare the different points of the model with various experimental analyses \cite{Drees:2013wra,Cao:2015ara} determining whether the point is excluded or not at 95\% C.L.. We applied the constraints on jets plus \met from ATLAS studies: ATLAS.1308.1841 \cite{Aad:2013wta}, ATLAS.1308.2631 \cite{Aad:2013ija}, ATLAS.1407.0583 \cite{Aad:2014kra} and ATLAS.1405.7875 \cite{Aad:2014wea}. In fact, we obtain that this point is indeed excluded by the analysis  ATLAS.1405.7875, in the signal region with 6 (or more) jets, with an $r \equiv \frac{S - 1.96 \Delta S}{S^{95}_{obs}}= 2.0$, where $S$ is the total number of expected signal events, $\Delta S$ is the total 1$\sigma$ uncertainty on this number and $S^{95}_{obs}$ is the experimentally measured 95\% confidence limit on signal events. 

Thus, we have to find another parameter point able to provide the signal, but satisfying all the present constraints on jets plus \met. The main difficulty in the GGM scenario is that each gluino decay produces typically four or more jets with large \pt and \met. Therefore, these points clash with constraints from observables with a large number of jets (plus \met), which have low backgrounds in the Standard Model. On the contrary, constraints with fewer jets are easier to satisfy because of larger backgrounds. Then, our strategy will be to compress sufficiently the spectrum to reduce the number of (observable) jets. This can be done by increasing the lightest neutralino mass closer to the gluino mass, so that the jets in the decays $\tilde g \to \chi_1 + {\rm jets}$ have smaller \pt.  With this goal, we construct a third GGM point.
The GGM3 point is obtained with $M_s = 4160$~TeV, $\tilde B_1^{1/2}=\tilde A_1=662.5$~TeV, $\tilde B_2^{1/2}=\tilde A_2 =344$~TeV, $\tilde B_3^{1/2}=117$~TeV, $\tilde A_3=330$~TeV and $\tan\beta=34.4$. With these parameters we obtain the spectrum shown in Table \ref{tab3}.
\begin{table}
\begin{center}
\begin{tabular*}{1.00\textwidth}{@{\extracolsep{\fill}}|c c c c c c c c c|}
\hline \hline
{\rm Particle}& $\tilde g$& $\chi_1^0$ & $\chi_2^0$ & $\chi_3^0$  & $\chi_4^0$ & $\chi_1^\pm$ & $\chi_2^\pm$& $\tilde G$~~~ \\
\hline
{\rm Mass (GeV)} & 985.4 & 905.7 & 918.7 & 1175.3 & 1190.6 & 910.2 & 1187.9 & $6.03 \times 10^{-7}$~~~ \\
\hline \hline
{\rm Particle}&  $\tilde q_L$ & $\tilde q_R$ & $\tilde b_1$ & $\tilde b_2$ & $\tilde t_1$ &  $\tilde t_2$ & $\tilde l_L$& $\tilde l_R$~~~ \\
\hline
{\rm Mass (GeV)} & 3140 & 2970 & 2820 & 2920 & 2630 & 2920 & 1330 & 1210~~~ \\
\hline\hline
{\rm Particle}&  $h$ & $H$ & $A$ & $H^+$ & & & & ~~~ \\
\hline
{\rm Mass (GeV)} & 121.0 & 1489 & 1489 & 1491 & & & & ~~~ \\
\hline
\end{tabular*}
\end{center}
\caption{\label{tab3} SUSY spectrum for the point GGM3.}
\end{table}
For this point, the BR$(\chi_1^0 \to \tilde G Z) = 0.98$ and the gluino production cross section is $(22.8 \pm 3.3)$ fb with Prospino at NLL+NLO. In this case, the signal is lower than in GGM2 but we still get $(13.1 \pm 2.2)$ events, well above the experimentally measured 95\% confidence limit on signal events.

Concerning the experimental searches on jets plus MET, according to Checkmate this point is allowed with an $r=1.0$ with the best signal region being 2 jets plus \met from ATLAS.1405.7875. Moreover, in the equivalent search  by CMS of two leptons, jets, and \met, CMS-SUS-14-014 \cite{Khachatryan:2015lwa}, we obtain $r=1.2$ in Checkmate. This result shows some tension with the ATLAS possitive signal from the bin with large \met in the CMS analysis, but still consistent with observations slightly above 95 \% C.L..

\begin{center}
  \begin{figure}[h!]
  \includegraphics[scale=0.5]{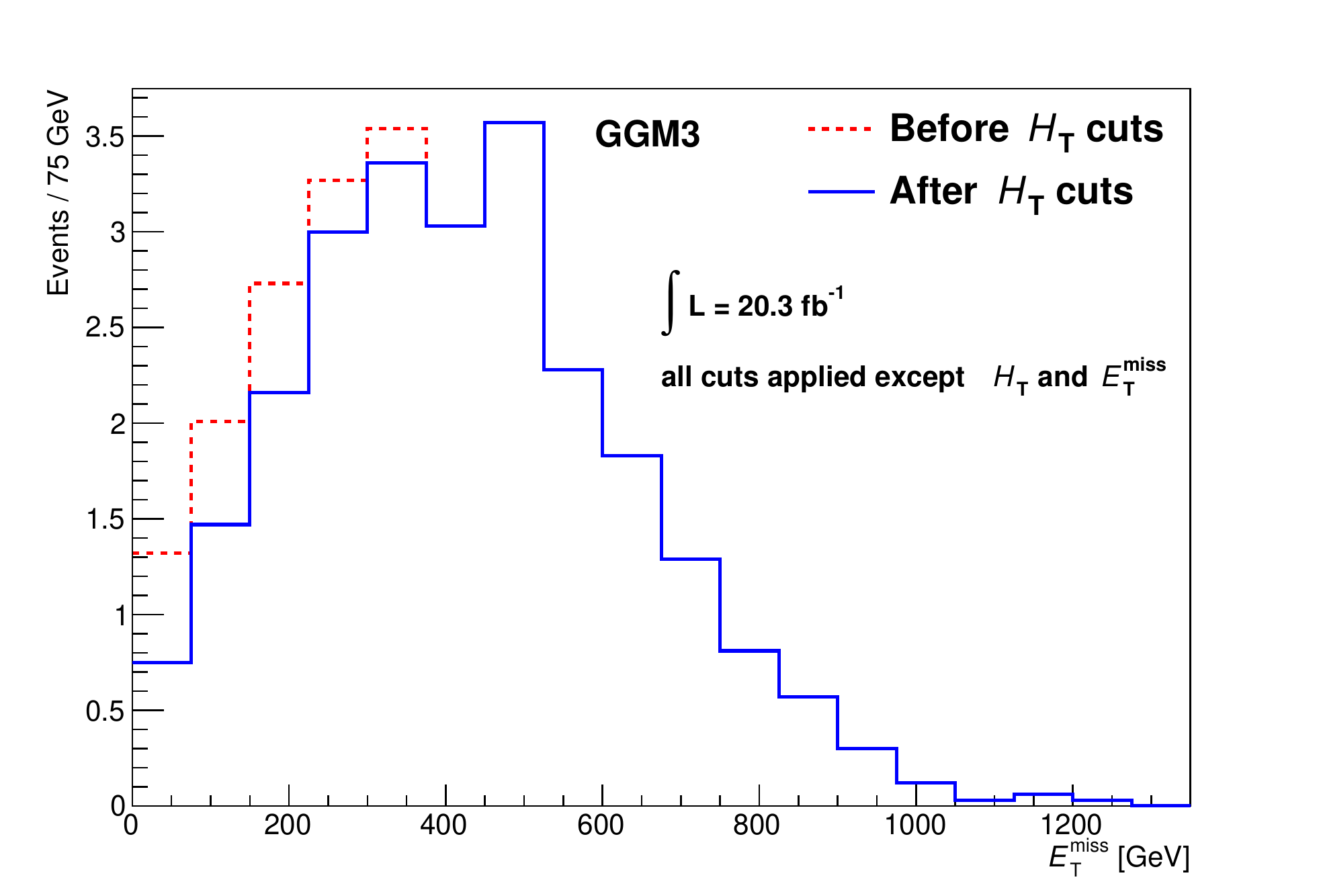} 
  \caption{\met distribution corresponding to the GGM3 point from strong production with the same selection as in Fig.~\ref{GGMHTMET}, before (red-dashed) and after (solid-blue) applying the $\HT>600~\GeV$ cut. Notice that electroweak production is negligible for the neutralino masses in GGM3.}
  \label{GGM3}
  \end{figure}
  \end{center}
  
Therefore, we have proved that it is indeed possible to construct a Supersymmetric model that accommodates the observed excess of lepton pairs on the $Z$-peak. The simulations presented here are only a proof of existence and the final model may be very different. Nevertheless, this model will have to share the main features of the examples that we presented here.

\section{Prospects for SUSY at LHC13}
As we have shown in the previous sections, the excess observed in ATLAS, if due to SUSY, requires a gluino of a mass $\sim 1$~TeV producing nearly one $Z$-boson per gluino in its decay. This scenario would also require relatively heavy squarks of the first generation with $m_{\tilde q} \gtrsim 2.5$~TeV.
If this is indeed the correct explanation to the observed excess, such light gluinos would be abundantly produced at Run II in LHC together with other SUSY particles.  Therefore, the results obtained at $pp$ collisions at 13 TeV will immediately confirm or reject this supersymmetric explanation of the ATLAS excess.
 
Using NLL-Fast with the spectrum of point GGM1, the gluino pair production cross-section at 8 TeV was $\sigma ( p p \to \tilde g \tilde g)_{\rm LHC8}^{\rm NLL+NLO} = 7.6 \pm 1.3$~fb at NLL+NLO.  Similarly the production cross-section at 13 TeV would be $\sigma ( p p \to \tilde g \tilde g)_{\rm LHC13}^{\rm NLL+NLO} = 150 \pm 16$~fb, that is, we would expect to produce 20 times more gluinos at LHC13 for the same integrated luminosity.   Repeating the same exercise with point GGM3, we have a gluino pair production cross-section at 8 TeV of $\sigma ( p p \to \tilde g \tilde g)_{\rm LHC8}^{\rm NLL+NLO} = 22.8 \pm 3.3$~fb while the production cross-section at 13 TeV would be $\sigma ( p p \to \tilde g \tilde g)_{\rm LHC13}^{\rm NLL+NLO} = 344 \pm 44$~fb. In this point, the cross section increases by a factor $\sim 15$ in going from 8 to 13~TeV center of mass energy. 
In any case, for both points this would have unambiguous signatures, both on the $Z$-peak with a scaling of the signal found at LHC8 and in direct searches for gluinos using jets plus missing $E_T$ in the extension of the analysis in \cite{Khachatryan:2015pwa,Aad:2015mia,Aad:2014wea}.

To confirm that it is indeed SUSY behind the excess found at the
$Z$-peak, we should search for other sparticles at LHC13. As shown in
Section~\ref{crosssection}, squarks of the first generation
are preferred be heavy to increase the gluino production
cross-section, namely above $2.5$~TeV. Then their production
cross-section at LHC13 would be around 1~fb which would make direct
detection very challenging. Charginos and neutralinos are expected to
be bellow  the gluino mass and in some cases could be abundantly
produced at LHC. Although our analysis does not fix the masses of
$\chi^\pm _1$ and $\chi^0_{1,2}$, they are expected to be rather heavy
$m_{\chi_1^0}\gtrsim 300$~GeV and degenerate. In some cases, the
production cross-section could be large. For instance $\sigma ( p p\to \tilde \chi^+_1 \chi^0_1)_{\rm LHC8}^{\rm NLO} = 60.2\pm 0.9$~fb
and $\sigma ( p p \to \tilde \chi^+_1 \chi^0_1)_{\rm LHC13}^{\rm NLO} = 154 \pm 5$~fb calculated with Prospino for $m_{\chi_1^0} = 313$ GeV and $m_{\chi_1^+} = 314$ GeV (with large $\tilde W$ component in $\chi^0_1$). Notice that,
as explained above, electroweak production is also large for these
points at 8 TeV, but is eliminated by the cuts on $\HT$ and number of
jets. Thus, a large electroweak production could be expected at LHC13
and dedicated searches should be encouraged, specially taking into
account the requirement of a large production of $Z$-bosons in
chargino and neutralino decays. Finally, the signal does not
constrain the masses of sleptons or third generation squarks and, a
priory, we can not make any prediction on their production at LHC13.

Before closing, we should comment on the nature of dark matter in our
scenario. As we have seen the signal seems to prefer a non-stable
neutralino decaying to a very light gravitino and a $Z$-boson. Under
these conditions the neutralino mass has no relation with the dark
matter abundance of the universe and its role as dark matter compon
ent would be played by the gravitino. The gravitino mass is not
bounded by the observed signal but regardless of its exact mass,
unfortunately no signal of dark matter is to be expected in direct
search experiments.

\section{Conclusions}

Recently the ATLAS experiment announced a 3~$\sigma$ excess at the $Z$-peak consistent of 29 pairs of leptons observed to be compared with $10.6 \pm 3.2$ expected lepton pairs. No excess outside the $Z$-peak was observed.
By trying to explain this signal with SUSY we found that only relatively light gluinos, $m_{\tilde g} \lesssim 1.2$~TeV, together with a heavy neutralino NLSP of $m_{\tilde \chi} \gtrsim 400$~GeV decaying predominantly to $Z$-boson plus a light gravitino, such that nearly every gluino produces at least one $Z$-boson in its decay chain, could do it. 

Unfortunately, this is not possible withing minimal SUSY models, as mSugra, minimal gauge mediation or anomaly mediation. The requirement of a neutralino NLSP decaying to $Z$ plus gravitino points to models of General Gauge mediation as the simplest possibility. We have shown that a model of this class is able to reproduce the observed signal overcoming all the experimental cuts. Needless to say, more sophisticated models could also reproduce the signal, however, they will ALWAYS share the above mentioned features, {\it i.e.} light gluinos (or heavy particles with a strong production cross-section) with an effective ${\cal N}(\tilde g \to Z) \simeq 1$.

\section*{Acknowledgments}
The authors are grateful to Francisco Campanario for useful discussions and Emma Torr\'o for her contributions in an early version of this paper. GB, JB and OV acknowledge support from the Ministry of Economy and Competitiveness (MINECO) and FEDER (EC)
Grants FPA2011-23596 and FPA2014-54459-P. GB, JB, VAM and OV from the Generalitat Valenciana under grant  PROMETEOII/2013/017. GB acknowledges partial support from the European Union FP7 ITN INVISIBLES (Marie Curie Actions, PITN- GA-2011- 289442). ER and VAM acknowledge support by MINECO under the project FPA2012-39055-C02-01.
VAM acknowledges support by the Spanish National Research Council (CSIC) under the JAE-Doc program co-funded by the European Social Fund (ESF).

\end{document}